\documentclass[aps, prd, showpacs, superscriptaddress, nofootinbib, twocolumn]{revtex4}
\usepackage{amssymb,amsmath,color,microtype}
\usepackage{graphicx}

\begin{document}
\title{New model of axion monodromy inflation and its cosmological implications}

\author{Yi-Fu Cai}
\email{yifucai@ustc.edu.cn}
\affiliation{CAS Key Laboratory for Researches in Galaxies and Cosmology, Department of Astronomy, University of Science and Technology of China, Chinese Academy of Sciences, Hefei, Anhui 230026, China}
\affiliation{Department of Physics, McGill University, Montr\'eal, QC, H3A 2T8, Canada}

\author{Fang Chen}
\email{fchen@kitp.ucsb.edu}
\affiliation{Kavli Institute for Theoretical Physics, University of California, Santa Barbara, California 93106, USA}

\author{Elisa G.~M.~Ferreira}
\email{elisafenu@physics.mcgill.ca}
\affiliation{Department of Physics, McGill University, Montr\'eal, QC, H3A 2T8, Canada}

\author{Jerome Quintin\footnote{Vanier Canada Graduate Scholar.}}
\email{jquintin@physics.mcgill.ca}
\affiliation{Department of Physics, McGill University, Montr\'eal, QC, H3A 2T8, Canada}

\begin{abstract}
We propose a new realization of axion monodromy inflation in which axion monodromy arises from torsional cycles in a type IIB compactification. A class of monomial potentials is obtained with specific values for the power index. Moreover, the inflaton mass changes profile due to the couplings between various fields after compactification. Consequently, the potential obtains a step-like profile at some critical scale. We study the cosmological implications of one concrete realization of this model. At the background level, it realizes a sufficiently long inflationary stage, which allows for the violation of the slow-roll conditions for a short period of time when the inflaton is close to the critical scale. Accordingly, the Hubble horizon is perturbed and affects the dynamics of primordial cosmological perturbations. In particular, we analyze the angular power spectrum of B-mode polarization and find a boost on very large scales. We also find that the amplitude of scalar perturbations is suppressed near the critical scale. Thus our model provides an interpretation for the low-$\ell$ suppression of temperature anisotropies in the CMB power spectrum. We examine these effects and confront the model to observations.
\end{abstract}


\maketitle

\section{Introduction}\label{sec:intro}

Large field inflationary models with super-Planckian field ranges are observationally allowed by the latest cosmological experiments~\cite{Ade:2013zuv, Planck:2013jfk}. The recent detection of primordial B-mode polarization reported by the BICEP2 experiment implies that the tensor-to-scalar ratio, $r$, which is the ratio between the amplitude of the spectra of tensor and scalar perturbations, is likely to be nonzero~\cite{Ade:2014xna}. This favors models of large field inflation via the Lyth bound~\cite{Lyth:1996im}. While the understanding of the BICEP2 data in terms of primordial tensor fluctuations becomes less clear~\cite{Mortonson:2014bja, Flauger:2014qra} because of a potential contamination by the dust foreground detected by the \textit{Planck} group~\cite{Adam:2014bub}, it remains of theoretical interests to study the realization of large field inflation within the framework of fundamental particle physics.

Considering the fact that inflation could occur near the Grand Unified Theory (GUT) scale, it is natural to ask whether large field inflation can be realized within string theory. In the literature, this topic has received extensive attention (for instance, see~\cite{Burgess:2013sla, Baumann:2014nda, Westphal:2014ana} for recent reviews). Recently, it was put forward and analyzed in~\cite{Silverstein:2008sg, McAllister:2008hb, Flauger:2009ab} that a string theory model of large field inflation can be achieved under the axion monodromy construction. In the corresponding stringy setup, a number of axion fields coupled to fluxes can yield a super-Planckian field variation while softly breaking the shift symmetry along the axion potential due to the coupling or due to D-branes. As a result, a class of axion monodromy inflation models with monomial potentials was obtained in~\cite{McAllister:2014mpa}. Various extended analyses of axion monodromy inflation were performed in the literature, for instance, in the context of super-gravity (SUGRA) realizations~\cite{Marchesano:2014mla, Blumenhagen:2014gta, Hebecker:2014eua}, by involving extra moduli fields~\cite{Ibanez:2014kia, Franco:2014hsa}, or by making connections to natural inflation~\cite{Freese:1990rb, Kim:2004rp, Kappl:2014lra, Long:2014dta, Bachlechner:2014hsa}. Also, a potential back-reaction issue against the moduli stabilization was recently discussed in~\cite{Blumenhagen:2014nba, Hayashi:2014aua, Hebecker:2014kva} from different perspectives.

In the present paper, we propose a new realization of axion monodromy inflation in terms of a set of D3- and D5-branes with
torsional cycles.
Our model represents a formulation of inflation from a class of string theory models presented in~\cite{Marchesano:2014mla}.
In our concrete construction, the potential for the inflaton field approximately takes the form of a power law function in which the exponent ($p$) varies depending on the detailed integral over the internal manifold as well as on the contribution from a Chern-Simons term. In general, these functions can appear as a linear combination in a single potential, which can be expressed as
$$V(\varphi) = \sum_i c_i \varphi^{p_i}~,$$
where the $c_i$'s are to be determined by the specific constructions. Therefore, our model can easily give rise to a sufficiently long period of inflation that is required by cosmological observations. More interestingly, we take a closer look at the shape of the inflaton's potential and we find that it may receive a modulation due to the uncertainty in the integration of the internal space. As a result, we expect this nontrivial modulation in the potential to leave important signals for cosmological experiments.

As a representative example, we study a specific inflation model derived from our construction. In particular, we analyze a model of chaotic inflation with a potential of the form $m^2\varphi^2$ in which the inflaton mass obtains a small modulation due to the variation of the internal space. Therefore, the inflaton potential possesses a step-like feature at a critical scale $\varphi_c$. We study in detail the cosmological implications of this model and we find that the mass modulation leads to a short phase of slow-roll violation during inflation when the inflaton field approaches $\varphi_c$. The Hubble horizon near the corresponding scale is also perturbed due to the violation of the slow-roll conditions. Accordingly, for the primordial fluctuations exiting the Hubble horizon at the same moment, the process of getting squeezed can be dramatically affected. In general, the spectrum of these fluctuation modes can become wiggly, but if the modulation is smooth enough, the spectrum can present a suppression behavior near the critical scale and then grow back to a larger amplitude at even larger length scales. We analyze these effects in the primordial angular power spectrum of temperature anisotropies and confront them with the \textit{Planck} 2013 data. Our results show that, for certain values of the model parameters, an appropriate suppression of the power can be achieved to explain the cosmic microwave background (CMB) anomaly at large angular scales. We further study the angular power spectrum of B-mode polarization that arises from primordial gravitational waves and obtain a small enhancement on very large scales. While being aware of the fact that the amplitude of primordial tensor fluctuations can always be modified by tuning the inflationary model parameters, the small amplification feature in the B-mode spectrum is a key prediction made by our model. This particular prediction may shed light on probing new physics by virtue of accumulating high-precision CMB polarization data in the near future.

The paper is organized as follows. In Sec.~\ref{sec:aminflation}, we first briefly review the general picture of axion monodromy inflation and then propose a new realization for axion monodromy inflation in string theory. Afterwards, we select a specific realization of our new axion monodromy model in Sec.~\ref{sec:data} and we study its cosmological implications in detail. We first study the background inflationary solution and analyze the dynamics of the slow-roll parameters. Then we numerically calculate the primordial power spectra of scalar and tensor perturbations generated during inflation. We then confront the theoretical predictions of our specific model with the latest CMB data by analyzing the temperature and B-mode polarization angular power spectra, and we perform Markov Chain Monte Carlo (MCMC) simulations to constrain the model. We conclude with a discussion in Sec.~\ref{sec:conclusion}. Throughout this paper, we adopt the convention $M_p^2 \equiv 1/8\pi G$.

\section{A model of axion monodromy inflation}\label{sec:aminflation}

In this section we first present a brief review of the regular version of axion monodromy inflation and then propose a new model in the presence of torsional cycles.

\subsection{A brief review of existing axion monodromy models}\label{sec:brief}

The idea of axion monodromy inflation originally proposed in~\cite{Silverstein:2008sg} has provided an interesting interpretation that a sufficiently long period of inflation can
persist through many cycles of the axion period while the inflaton's potential is well
controlled by the axion shift symmetry. This model, and related extensions, were widely
studied since then (see, e.g.,~\cite{McAllister:2008hb, Flauger:2009ab, Berg:2009tg, Hannestad:2009yx, Conlon:2011qp, Shlaer:2012by, Marchesano:2014mla, Gao:2014uha, Ozsoy:2014sba}
and references therein).

In most of these models, the axion field comes from the reduction of a NS-NS two-form $B_2$
on a two-cycle which has a continuous shift symmetry due to the higher dimensional gauge
symmetry of the two-form, and the monodromy is induced by branes or fluxes.  As a simple
example, one can consider a D5-brane wrapping an internal two-cycle $\Sigma_2$ in type IIB string
theory. The axion, which is given by $b=\frac{1}{\alpha'}\int_{\Sigma_2} B_2$, has a
shift symmetry $b\rightarrow b+\mathrm{const}$. However, the presence of the D5-brane gently breaks
such a shift symmetry and generates the monodromy. To be explicit, we can consider
the Dirac-Born-Infeld (DBI) action for the D5-brane,
\begin{eqnarray}
 S_{D5}=\frac{1}{(2\pi)^5g_s\alpha^{'3}}\int_{\mathcal{M}_4\times \Sigma_2} d^6x~\sqrt{-\mathrm{det}(G_{ab}+B_{ab})} ~,
\end{eqnarray}
which is an effective field description of string theory at low energy scales.
Here, $\mathcal{M}_4$ denotes the $3+1$ space-time, and $G_{ab}$ and $B_{ab}$, where $a,b=0,...,5$, are the induced metric
and NS-NS two-form on the D5-brane world volume, respectively.

After integrating over the two-cycle $\Sigma_2$, one obtains the potential of the axion
in the four-dimensional effective theory,
\begin{eqnarray}
 V(b)=\frac{\rho}{(2\pi)^6g_s\alpha^{'^2}}\sqrt{(2\pi)^2\ell^4+b^2} ~,
\end{eqnarray}
where $\ell$ is the size of the two-cycle $\Sigma_2$ in string units and $\rho$ is a
dimensionless coefficient determined by the warp factor.  For large values of $b$, the
potential is linear, $V(b)\propto b$. Correspondingly, the Lagrangian of the canonically
normalized inflaton field $\varphi$ is given by,
\begin{eqnarray}
 \mathcal{L}=-\frac{1}{2}(\partial \varphi)^2-m^3 \varphi, \,\,\, {\rm{with} }\,\,\,
 m^3=\frac{1}{f}\frac{\rho}{(2\pi)^6 g_s\alpha^{'^2}}~,
\end{eqnarray}
where $f$ is the axion decay constant. In addition, a similar scenario can be achieved
when the D5-brane is replaced by an NS5-brane.

It has been interestingly observed that there are many variants of the axion's potential
that may arise from axion monodromy. Thus, it is convenient to parameterize these theories
in terms of an exponent $p$:
\begin{eqnarray}
 \mathcal{L}=-\frac{1}{2}(\partial \varphi)^2-m^{4-p}\varphi^p ~.
\end{eqnarray}
For example, a model with $p=2/3$ was obtained in~\cite{Silverstein:2008sg} based on a
construction of Nil manifolds; the case of $p=2$ was found after taking into account an
appropriate coupling of the axion to a four-form field~\cite{Kaloper:2008fb}; more
examples with $p=2/3$, $4/3$, $2$ and $3$ were obtained in~\cite{McAllister:2014mpa} by
considering effective couplings of the NS-NS $B_2$ field to the RR $F_1$ field through
$|\tilde{F}_5|^2$ and $|\tilde{F}_3|^2$, respectively, after reducing type IIB on a circle with specific
background fluxes and moduli stabilization.

Another interesting model of axion monodromy inflation was constructed in
\cite{Marchesano:2014mla}, where the axion arises from the KK compactification of higher
dimensional gauge fields or p-form potentials in the presence of fluxes and/or torsion
homology. The presence of these extra components generate the F-term potential.
In the same work the authors argued that this scenario could naturally avoid the $\eta$-problem widely existing in the inflationary paradigm~\cite{Copeland:1994vg} since those axions did not appear in the K\"{a}hler potential as they did in other models.

\subsection{New axion monodromy inflation from warped torsion classes}\label{sec:newaxion}

We propose a new model of axion monodromy inflation in the present subsection.
This represents a realization from a class of axion monodromy models where the
F-term potential comes from the inclusion of extra ingredients, as seen in~\cite{Marchesano:2014mla}.
In our model we include torsional cycles.
Specifically, we consider a type IIB string theory compactification and assume that there are
$N$ D3-branes along the $3+1$ space-time. Thus the metric naturally has a warp factor
$e^{-2A}\sim 1/N$, maintaining the four-dimensional Poincar\'e invariance, and takes the form
\begin{eqnarray}
 ds_{10}^2=e^{-2A(y^i)}dx_{\mu}dx^{\mu}+e^{2A(y^i)}dy_idy^i~,
\label{eq:metric}
\end{eqnarray}
where $\mu=1,...,4$ represents the coordinates of the physical space-time, and $i=1,...,6$
runs over the coordinates of the internal manifold $\mathcal{M}_6$.

It is well known that in a (conformal) Calabi-Yau (CY) compactification, the
massless modes are in one-to-one correspondence with elements of the cohomology groups of the
internal manifold, modulo certain subtleties that arise once one introduces warping
\cite{DeWolfe:2002nn,deAlwis:2003sn,Giddings:2005ff,Shiu:2008ry,Douglas:2008jx,Frey:2008xw,Underwood:2010pm}.
However, the situation becomes slightly different
if there are torsional cycles since in this case, the massless modes are not sensitive
to D-branes wrapping torsional cycles. In the following we use $X_6$ to denote the unwarped
internal manifold and distinguish it from $\mathcal{M}_6$. In general, there are two
independent torsion classes in a six-dimensional manifold, i.e.,  torsion one-cycles
$\Sigma_1$ and torsion two-cycles $\Sigma_2$. They can be related to torsion four-cycles
$\Sigma_4$ and torsion three-cycles $\Sigma_3$ via
\begin{eqnarray}
 \mathrm{Tor}H_r(X_d,\mathbb{Z})=\mathrm{Tor}H_{d-r-1}(X_d,\mathbb{Z})~,
\end{eqnarray}
for a $d$-dimensional internal manifold and any positive integer $r$.
For simplicity we take
\begin{align}
 & { \rm{Tor}} H_1(X_6,\mathbb{Z})={\rm{Tor} }H_4(X_6,\mathbb{Z})=\mathbb{Z}_p \nonumber\\
 & { \rm{Tor}} H_2(X_6,\mathbb{Z})={\rm{Tor} }H_3(X_6,\mathbb{Z})=\mathbb{Z}_k ~,
\end{align}
where $p$ and $k$ are positive integers and in general not equal. Now let us consider a set
of non-harmonic Laplacian eigenforms of $X_6$,
\begin{align}
 & \mathrm{d}\gamma_1=p\rho_2 ~,\; \mathrm{d}\tilde{\rho}_4=p\tilde{\gamma}_5~, \\
 & \mathrm{d}\eta_2=k \omega_3 ~,\; \mathrm{d}\tilde{\omega}_3=k\tilde{\eta}_4~,
\end{align}
where $\gamma_1,\ \tilde{\rho}_4,\ \eta_2$, and $\tilde{\omega}_3$ are associated with the
generators of ${ \rm{Tor}} H_i(X_6,\mathbb{Z})$ with $i=1$, $4$, $2$, and $3$, respectively. We note
that, $\rho_2,\ \tilde{\gamma}_5,\ \omega_3$, and $\tilde{\eta}_4$ are trivial in de Rham
cohomology but are non-trivial generators of $H^{i}(X_6,\mathbb{Z})$, $i=2,5,3,4$. Additionally,
we have the following integral constraints:
\begin{align}
 \int\gamma_1\wedge \tilde{\gamma}_5 = \int \rho_2\wedge \tilde{\rho}_4 = \int\eta_2\wedge \tilde{\eta}_4 = \int \tilde{\omega}_3\wedge\omega_3 = 1 ~.
\end{align}
Then we can expand type IIB NS-NS and RR forms in terms of these eigenforms as
\begin{align}\label{expand}
 B_2 = & b\eta_2+\bar{b}\rho_2+b_1\wedge \gamma_1+b_2~,\nonumber\\
 C_2 = & c\eta_2+\bar{c}\rho_2+c_1\wedge \gamma_1+c_2~,\nonumber\\
 C_4 = & \mathfrak{c}\tilde{\rho}_4+\bar{\mathfrak{c}}\tilde{\eta}_4+\mathfrak{c}_1\wedge\tilde{\omega}_3+\bar{\mathfrak{c}}_1\wedge \omega_3\nonumber\\
 & +\mathfrak{c}_2\wedge\rho_2+\bar{\mathfrak{c}_2}\wedge\eta_2+\mathfrak{c}_3\wedge\gamma_1+\mathfrak{c}_4 ~,
 \end{align}
and they are the most general expansions regardless of the orientation.
The corresponding ten-dimensional field strengths are given by
\begin{align}\label{expand2}
 H_3 \equiv \mathrm{d}B_2
 =& \mathrm{d}b\wedge \mathrm{d}\eta_2+kb\omega_3+(\mathrm{d}\bar{b}-pb_1)\wedge\rho_2\nonumber\\
&+\mathrm{d}b_1\wedge\gamma_1+\mathrm{d}b_2~,\nonumber\\
 F_3\equiv \mathrm{d}C_2
=&\mathrm{d}c\wedge \mathrm{d}\eta_2+kc\omega_3+(\mathrm{d}\bar{c}-pc_1)\wedge\rho_2\nonumber\\
&+\mathrm{d}c_1\wedge\gamma_1+\mathrm{d}c_2~,\nonumber\\
 F_5 \equiv \mathrm{d}C_4
=&\mathrm{d} \mathfrak{c}\wedge\tilde{\rho}_4+p\mathfrak{c}\tilde{\gamma}_5+(\mathrm{d}\bar{\mathfrak{c}}-k\mathfrak{c}_1)\wedge\tilde{\eta}_4 +\mathrm{d}\mathfrak{c}_1\wedge\tilde{\omega}_3~\nonumber\\
 &+(\mathrm{d}\bar{\mathfrak{c}}_1+k\bar{\mathfrak{c}}_2)\wedge \omega_3+(\mathrm{d}\mathfrak{c}_2-p\mathfrak{c}_3)\wedge\rho_2~\nonumber\\
 &+\mathrm{d}\bar{\mathfrak{c}}_2\wedge\eta_2+\mathrm{d}\mathfrak{c}_3\wedge\gamma_1+\mathrm{d}\mathfrak{c}_4 ~.
\end{align}

Next, we perform the dimensional reduction of the type IIB SUGRA action on
$\mathcal{M}_6$. The type IIB SUGRA action including local sources in ten dimensions is then given by
\begin{align}\label{action}
 S = & \frac{1}{2\kappa^2_{10}}\int d^{10}x\,\sqrt{G}\Big(R+\frac{\partial_M\tau\partial^M\bar{\tau}}{2|\rm{Im}\tau|^2}-\frac{|\tilde{F}_5|^2}{4\cdot 5!}-\frac{G_3\cdot\bar{G}_3}{12\rm{Im}\tau}\Big)\nonumber\\
 &+\frac{1}{8i\kappa_{10}^2}\int\frac{C_4\wedge G_3\wedge \bar{G}_3}{\rm{Im}\tau}+S_{\mathrm{loc}}
\end{align}
in Einstein frame, where $\tau=C_0+ie^{-\phi}$ is the axion-dilaton field with $C_0$
being the string axion (one should be aware that this is not the same axion field
as the one we will discuss below in the four-dimensional effective theory) and $\phi$ being the dilation
field. Moreover,
\begin{equation}
 \tilde{F_5}=\mathrm{d}C_4-\frac{1}{2}C_2\wedge H_3+\frac{1}{2}B_2\wedge F_3~,
\end{equation}
which is the five-form flux, and
\begin{equation}
 G_3=F_3-\tau H_3~,
\end{equation}
with $F_3$ being the RR three-form flux
and $H_3$ the NS-NS three-form flux. Also, $G$ is the determinant of the ten-dimensional
metric $g_{MN}$, where $M,N=0,...,9$. Finally, $S_{\mathrm{loc}}$  is the action for localized
sources in the system, i.e. D3-branes in our case.

The presence of this finite number of localized D3-branes leads to the appearance of a D3-brane charge density.
Integrating the Bianchi identity (or equation of motion) for the five-form flux over $X_6$,
we arrive at the tadpole cancellation condition,
\begin{align}
Q_3^{\mathrm{loc}}+\frac{1}{2\kappa ^2 T_3}\int _{X_6} H_3 \wedge F_3 =0~,
\label{tadpole}
\end{align}
where $Q_3^{\mathrm{loc}}$ is the charge associated with the D3-brane.
The charge must be conserved so that the tadpole condition is not violated.
Here, the presence of the $N$ D3-branes leads to the presence of D3-brane charges since at every axion period
the B-fields induce these charges.
To solve this problem, we propose the inclusion of $3$-form fluxes.
Those fluxes carry the amount of D3-brane charge given by the integral in\ \eqref{tadpole}.
So, by a suitable choice of fluxes, the tadpole condition can be met.
Another consequence is that their inclusion is responsible for stabilizing the complex structure and the dilaton moduli.
We comment in the next subsection in more detail about the inclusion of those fluxes and about the moduli stabilization issue.

Plugging Eq.~\eqref{expand} and Eq.~\eqref{expand2} into Eq.~\eqref{action}
gives rise to the four-dimensional effective action that we are interested in.
In particular, we are mostly interested in the contributions from the scalar fields
$b$, $c$ and $\mathfrak{c}$, and hence, the effective action in four dimensions can
be simplified as
\begin{align}\label{S_4_bcc}
 S_4 =& -\int d^4x~\frac{\sqrt{g_4}}{24\kappa_{10}^2}\Big[(\mathcal{T}_1+\mathcal{T}_2c^2)\partial_{\mu} b\partial^{\mu}b\nonumber\\
 &+(\mathcal{T}_1+\mathcal{T}_2b^2)\partial_{\mu} c\partial^{\mu}c+k^2\mathcal{T}_3(b^2+c^2) \nonumber\\
 &+\mathcal{T}_2bc\,\partial_{\mu}b\partial^{\mu}c  +\mathcal{T}_4\partial_{\mu}\mathfrak{c}\partial^{\mu}\mathfrak{c} +p^2\mathcal{T}_5\mathfrak{c}^2\Big] \nonumber\\
 & + ...~,
\end{align}
where
\begin{align}
\mathcal{T}_1=&\int\eta_2\wedge \star_6\eta_2,\;\;\;\mathcal{T}_2=\int e^{-4A}\,\, \eta_2\wedge \eta_2\wedge\star_6(\eta_2\wedge \eta_2),\nonumber\\
\mathcal{T}_3=&\int e^{-4A}\,\, \omega_3\wedge\star_6\omega,\;\;\;
\mathcal{T}_4=\int e^{-4A} \,\, \tilde{\rho}_4\wedge\star_6\tilde{\rho}_4,\nonumber\\
\mathcal{T}_5=&\int e^{-8A} \,\, \tilde{\gamma}_5\wedge\star_6\tilde{\gamma}_5.
\end{align}
Also, $g_4$ is the determinant of the unwarped space-time metric and the hodge star
$\star_6$ is associated with the unwarped internal manifold $X_6$.
In the above, we have set the axion $C_0$ and dilaton $\phi$ to zero.
We also work only at leading order in $\alpha'$ ; at higher orders in
$\alpha'$, there will perturbative corrections to the kinetic terms,
as studied for example in \cite{Anguelova:2010ed}, in addition to the usual curvature
corrections\footnote{As mentioned below equation\ \eqref{eq:metric}, there are also
subtleties that arise once one considers non-small warping. In
particular, solving the Einstein equation requires a careful analysis
of the `breathing mode' and the introduction of a compensator field
(see, e.g.,\ \cite{Underwood:2010pm}).}.
The ellipsis at the end of the action denotes the contributions from the ten-dimensional Chern-Simons term
and other relevant terms, which are all linear in the $b$, $c$, and $\mathfrak{c}$ fields.

To normalize the $b$ field, we introduce
\begin{equation}
 \mathfrak{b}=\frac{\kappa_4\sqrt{\mathcal{T}_1+\mathcal{T}_2c^2}}{2\sqrt{3}\kappa_{10}}\,b~.
\end{equation}
Thus the $b$- and $c$-related terms in the action can be written as
\begin{align}
 S_4 \sim& -\int d^4x~\frac{\sqrt{g_4}}{2\kappa_{4}^2}\Big[\partial_{\mu} \mathfrak{b}\partial^{\mu}\mathfrak{b}+\frac{k^2\mathcal{T}_3}{\mathcal{T}_1+\mathcal{T}_2c^2}\mathfrak{b}^2\nonumber\\
 &+(\mathcal{T}_1+\mathcal{T}_2b^2(\mathfrak{b}))\partial_{\mu} \tilde{c}\partial^{\mu}\tilde{c}+k^2\mathcal{T}_3\tilde{c}^2 \Big]~,
\end{align}
where $\tilde{c}=\kappa_4c/(2\sqrt{3}\kappa_{10})$.

The above three scalar fields $\mathfrak{b}$, $c$ (or $\tilde{c}$ equivalently), and $\mathfrak{c}$
can be candidates for the inflation field. On one hand,
as has been argued in~\cite{Marchesano:2014mla}, in the (conformal)
CY compactification the $b$ field is generally involved in the K\"{a}hler potential.
Thus, if one takes the $\mathfrak{b}$ field as the inflaton field,
upon the moduli stabilization, the corresponding potential could  achieve a correction that
renders a possibly large value of the $\eta$ parameter ($\eta\sim1$), and therefore may
invalidate the occurrence of inflation. On the other hand, the $\mathfrak{c}$ field gives rise to the
standard chaotic inflation by checking the last two terms of the action \eqref{S_4_bcc}.
We also note that the $\mathfrak{c}$ field is decoupled from the other two fields.
Therefore, we focus on the cosmological implications of the $\tilde{c}$ field.

Before we study the inflaton field, we also have to stabilize the $\mathfrak{b}$ field.
We notice that the $\mathfrak{b}$ field couples to the $c$ field in a non-standard
way and that the value $\mathfrak{b}_{\mathrm{min}}$ where its potential is the lowest depends on $c$.
Since we know that $e^{-4A}\sim 1/N^2$, the $c$ dependence is mild if $c< N^2$ and strong if $c> N^2$.
In particular, for $c< N^2$ we ignore its dependence in the $\mathfrak{b}$ potential and $\mathfrak{b}_{\mathrm{min}}$ is fixed; for $c>N^2$ we will see that it can give rise to cosmological inflation.
Thus, we can set $c=N^2$ temporarily in order to find $\mathfrak{b}_{\mathrm{min}}$ in this region.

Now that $\mathfrak{b}_{\mathrm{min}}$ is fixed, $b_{\mathrm{min}}$ can also be determined.
Thus we can normalize the $c$ field and
write down the four-dimensional effective action as follows,
\begin{eqnarray}\label{action_phi_4D}
 S_4=-\frac{1}{2}\int d^4x~\sqrt{g_4} \Big( \partial_{\mu}\varphi\partial^{\mu}\varphi +m^2\varphi^2 \Big) ~,
\end{eqnarray}
where
\begin{align}
 \varphi &=\sqrt{\mathcal{T}_1+\mathcal{T}_2b_{\mathrm{min}}^2}\,\tilde{c}~, \nonumber\\
 m^2 &=\frac{k^2}{\mathcal{T}_1+\mathcal{T}_2b_{\mathrm{min}}^2}~.
\end{align}

One could do another field transformation in Eq.~\eqref{action_phi_4D} to absorb the linear term
and leave the mass $m^2$ unchanged. The resulting cosmological solution would lead to simple chaotic inflation.
Instead, we would like to explore another possibility in which the mass of the inflaton field
varies along the inflaton value, which occurs at
\begin{equation}
 \varphi_{c}=\frac{\kappa_4\sqrt{\mathcal{T}_1+\mathcal{T}_2b_{\mathrm{min}}^2(N^2)}\,N^2}{2\sqrt{3}\kappa_{10}}~.
\end{equation}
This can be achieved when the flux coupling, which determines the mass of the inflaton field,
varies between the inside and outside of the warped manifold.
Such a variation of the flux coupling can lead to a modulation at a critical scale.
To see this more explicitly, we rewrite the mass term as follows,
\begin{align}
 m^2 \varphi^2&=\Big[m_a^2+(m_b^2-m_a^2)\theta(\varphi-\varphi_{c})\Big]\varphi^2~,
\end{align}
where $m_a^2$ and $m_b^2$ are the inflaton masses when the inflaton field is smaller and larger
than the critical value $\varphi_{c}$, respectively.
We have used the $\theta$ function here to describe the most extremal case. However, as we discussed previously,
$b_{\mathrm{min}}(c)$ is smooth, so instead of the step function $\theta$, it is better to use the following parametrization,
\begin{align}
 m^2 \varphi^2&=m_a^2 \varphi^2 + \frac{(m_b^2-m_a^2) \varphi^2 }{1+\exp\left[-2 C_H (\varphi^2-\varphi_c^2)/M_p^2\right]}~.
\end{align}
The parameter $C_H$ depends on the smoothness of $b_{\mathrm{min}}(c)$ and thus on the potential of the $\mathfrak{b}$ field.
From now on we will treat it as a free parameter, but since no transition can be super-Planckian in string theory,
we do not expect it to be larger than 1.

\subsubsection{General details of the construction}

Without the presence of fluxes in our theory there are massless fields corresponding to the complex structure moduli,
dilaton modulus, and K\"{a}hler moduli. The tadpole condition is also not obeyed since D3-brane sources were included.
We now propose a procedure\ \textit{\`a la} KKLT~\cite{Kachru:2003aw,Giddings:2001yu,Baumann:2014nda} in order to stabilize the
complex structure, dilaton, and K\"{a}hler moduli in a two-step procedure. We uplift the solution to a de Sitter solution as an
extra final step. Namely, we propose to include suitable $3$-form fluxes in order to stabilize the complex structure $z$
and dilaton moduli $\tau$, and also meet the tadpole cancellation condition. Following that, we introduce non-perturbative
corrections to stabilize all K\"{a}hler moduli, $\rho$, and all the others coming from cycles. It is not in the scope of this paper to work out these calculations in
detail, but we wish to give a general view of our construction.

We want to include fluxes in our theory. The way these fluxes can be included in our compactified construction without
influencing our background is by gluing a non-compact manifold to our compactified CY, similar to the simple mechanism done
in~\cite{Giddings:2001yu}, where the fluxes are included in a conifold of the CY manifold (an extension of the Klebanov-Strassler
work for compact manifolds)~\cite{Baumann:2014nda}. We assume that the internal space is very large, in a way that supergravity
is still valid, but with a size where the $4$-dimensional Newton's constant is still finite (non-vanishing), and that the non-compact manifold is located far away from the warped region where our model is described.
So, we include localized fluxes in the non-compact manifold far away from where we added our sources in the compactified
manifold. We assume that there is no backreaction, which means that the fluxes are not coupled to the sources and that their
influence dies off towards the direction of the D3-branes. In this way, the physics in the neighborhood of the D3-branes
and the torsional cycles is essentially unchanged.

The presence of these localized $3$-form fluxes in the non-compact manifold does not change locally what is happening
where our theory lives. However, it changes globally the theory. The tadpole condition, or charge conservation, is a global
condition. So, no matter where one localizes those fluxes, only the presence of fluxes with negative D3-brane charges is enough
to ensure that the tadpole condition is met. The complex structure and dilaton moduli are stabilized by the presence of those
fluxes. The fluxes generate a Gukov-Vafa-Witten (GVW) flux superpotential ($W$), which stabilizes the complex structure and dilaton moduli at a
supersymmetric minimum of the potential, $D_{\tau}W=D_z W=0$. So, the presence of fluxes fixes all of the complex structure
and the dilaton moduli but leaves all the K\"{a}hler moduli unfixed. We need to tune those fluxes correctly in order to precisely
obey the tadpole condition and control those moduli. Working out explicitly these solution is beyond the scope of this paper.

Now, we have to stabilize the K\"{a}hler modulus~\cite{Kachru:2003aw,Giddings:2001yu,Baumann:2014nda}. This can be done by
inducing a potential via a non-perturbative effect like gaugino condensation on a D7-brane or if the four-cycle $\Sigma_4$
is wrapped by D3-brane instantons, also called Euclidean D3-branes. By finding the minimum at $D_{\rho}W=0$, the K\"{a}hler
modulus is stabilized.

The vacuum found above has an AdS minimum with negative energy. Since we are interested in inflationary solutions, we need to
uplifting the AdS minimum to a de Sitter vacuum. This can be done by any of the known mechanisms in the
literature, for example~\cite{Kachru:2003aw,Baumann:2014nda}. A different approach can be found in\ \cite{Dasgupta:2014pma}. The KKLT procedure is to add an uplifting potential to the theory, which
breaks SUSY, and thus slightly shifts the minimum without disrupting the physics that led to the stabilized minimum.
More details on this procedure can be seen in~\cite{Baumann:2014nda}.

The above details give a broad view of how to realize our new axion monodromy model. The resolution of some of those topics is still
unknown in the field, and these problems are present in many string inflation models. Those are material for deeper studies in axion
monodromy and in general string inflation.

Also, in the present study, we have adopted a much simplified treatment of the warping effects in deriving the four-dimensional
effective action. In the literature, there are a number of efforts on this subject, for example,
see\ \cite{DeWolfe:2002nn,deAlwis:2003sn,Giddings:2005ff,Shiu:2008ry,Douglas:2008jx,Frey:2008xw,Underwood:2010pm}
for relevant extensive studies and\ \cite{Douglas:2006es,Denef:2007pq} for comprehensive reviews on the topic of flux compactification.
Since the major focus of this paper is the cosmological implications of our model,
we wish to leave more detailed studies of this aspect for future work.

\section{Inflationary dynamics and cosmological implications}\label{sec:data}

In this section, we select one specific model of the new axion monodromy model as a
demonstration to study its cosmological implications. Specifically, we consider the effective action in \eqref{action_phi_4D} and assume a negligible Chern-Simons term. Then the Lagrangian derived at the end of the previous section
can be rewritten as
\begin{equation}
 \label{L_inflation}
 \mathcal{L}=-\frac{1}{2}\partial_\mu\varphi\partial^\mu\varphi-V(\varphi)~,
\end{equation}
which is viewed as an effective Lagrangian in 4-dimensional spacetime. The potential of
the inflaton takes the form of $m_b^2\varphi^2/2$ when $|\varphi| > \varphi_c$
(which corresponds to the UV regime) while taking the form of $m_a^2\varphi^2/2$ when
$|\varphi| \leq \varphi_c$ (which belongs to the IR regime). We note that $m_b>m_a$
is required for the model to be consistent with theoretical constraints. For convenience,
we would like to parameterize the potential in the following form,
\begin{align}\label{V_para}
 V(\varphi) = \frac{1}{2}m_a^2 \varphi^2 + \frac{(m_b^2-m_a^2) \varphi^2 }{2\left\{1+\exp\left[-2 C_H (\varphi^2-\varphi_c^2)/M_p^2\right]\right\}} ~,
\end{align}
where we have introduced one dimensionless coefficient $C_H$ in order to smoothly connect
the UV and IR regimes of the potential near the critical scale characterized by
$\varphi_c$. This newly introduced parameter ought to be determined by the
explicit physics of the mass transition in the corresponding string theory construction.
However, due to the lack of specific stringy derivations, we argue that $C_H$
can be treated as a free parameter in our model that can be constrained by cosmological
observations as will be analyzed in the following subsections.

\begin{figure}
\includegraphics[scale=0.3]{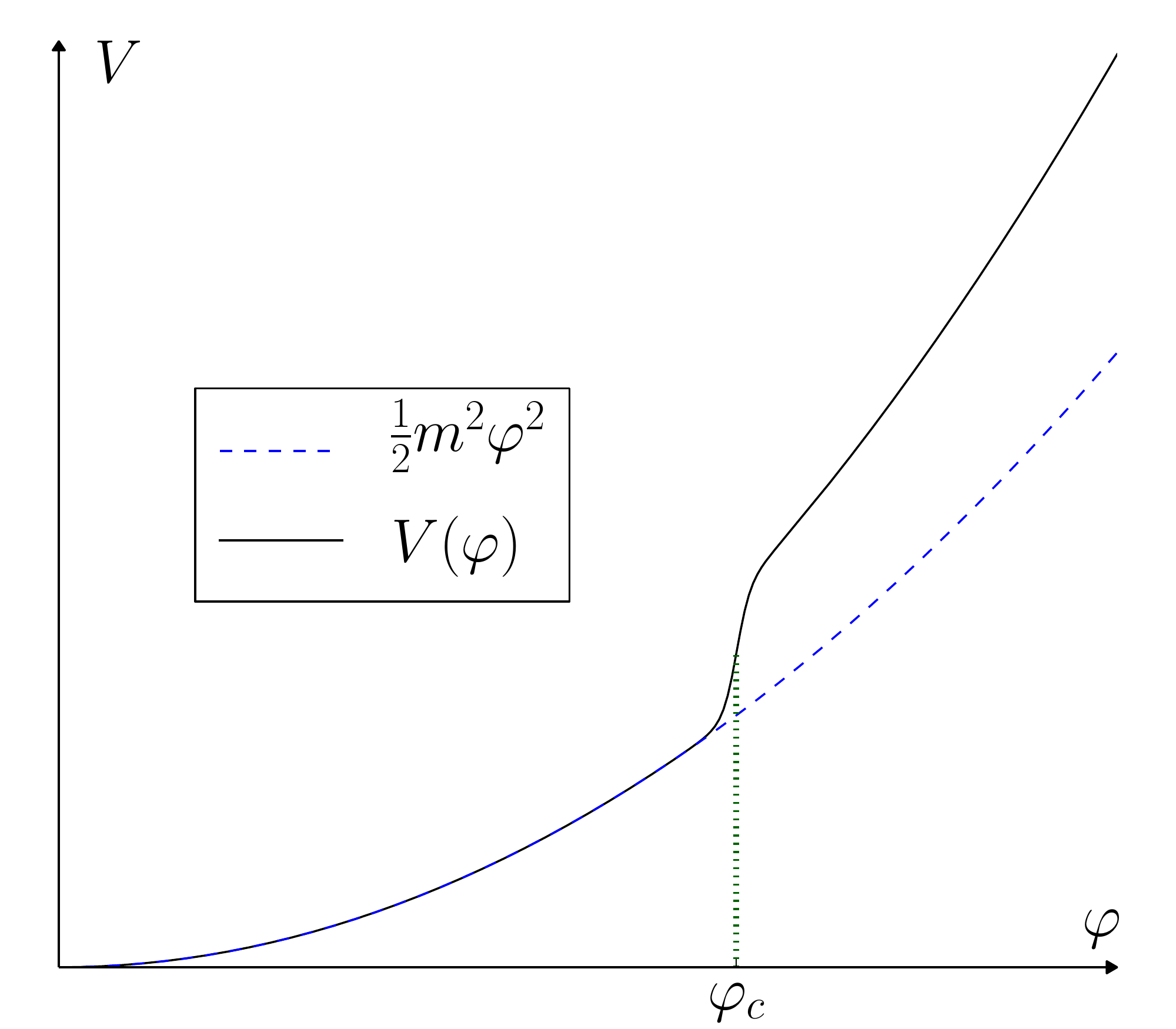}
\caption{Sketch of the potential $V$ as a function of the inflation scalar field
$\varphi$ for our model (the solid black line) and for regular chaotic inflation (the
dashed blue line). The potential under consideration takes the form of a quadratic
function at both high and low energy scales, but there is a step-like profile that
connects these two regimes at a critical scale $\varphi_c$. From the perspective of
string theory, it indicates a vibration of the internal space. In our specific
parametrization, we smooth out the step profile, the steepness of which is characterized by
the dimensionless parameter $C_H$.}
\label{Fig:sketch}
\end{figure}

We provide a sketch of the parameterized potential $V(\varphi)$ given by Eq.~\eqref{V_para} in Fig.~\ref{Fig:sketch}. This figure is helpful to gain a semi-analytic understanding of the background dynamics, which is the topic we turn to in the following subsection. We note that features in the inflationary potential such as above have a long history, namely see~\cite{Starobinsky:1992ts} for a pioneer study and~\cite{Adams:1997de, Adams:2001vc} for phenomenological constructions. Also, we note that a similarly featured potential was implemented phenomenologically by
a model of multi-field inflation in~\cite{Feng:2003zua, Joy:2007na, Joy:2008qd}.

\subsection{Background dynamics}

\begin{figure*}
\begin{center}
\includegraphics[scale=0.48]{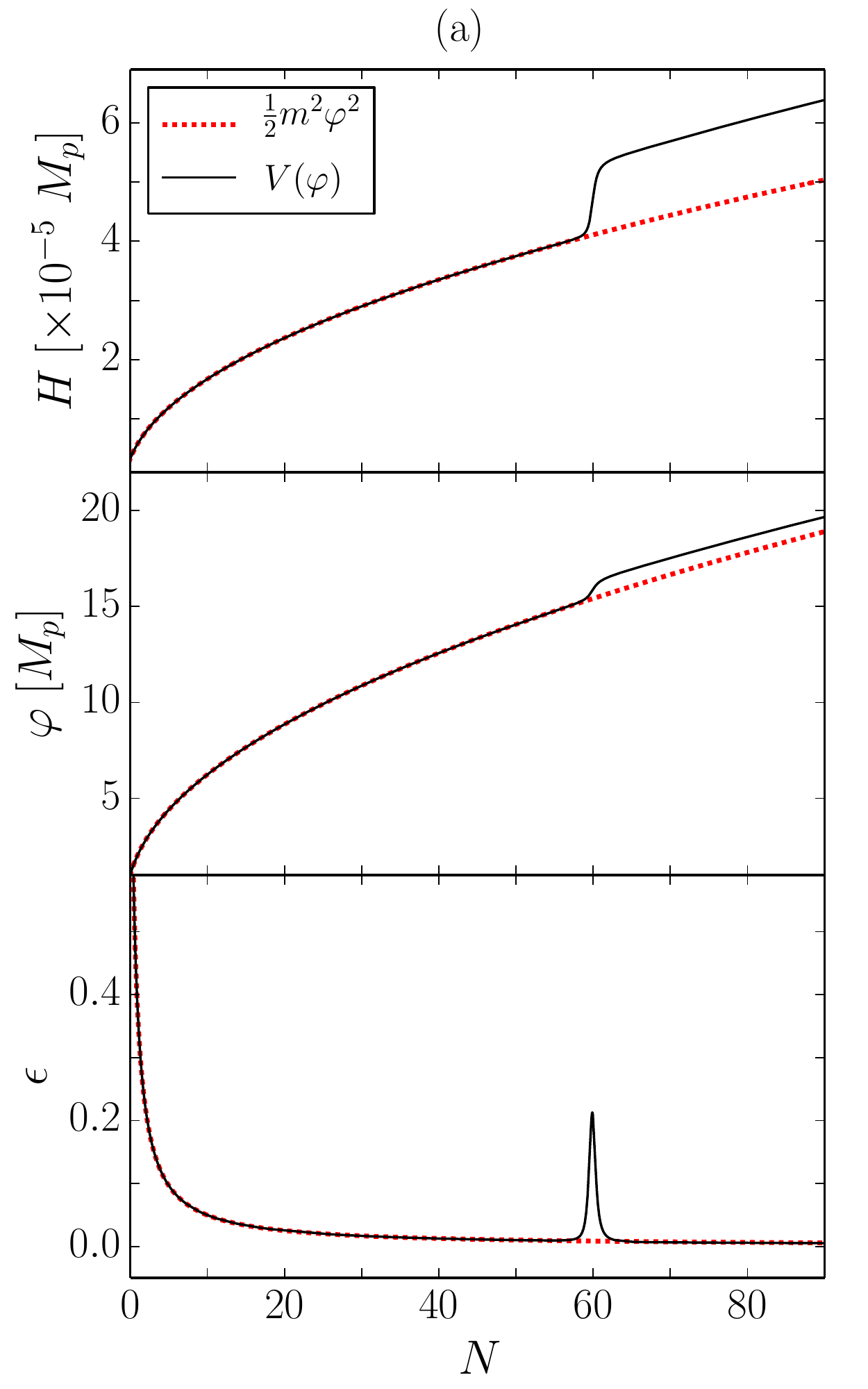} \hglue0.2cm
\includegraphics[scale=0.48]{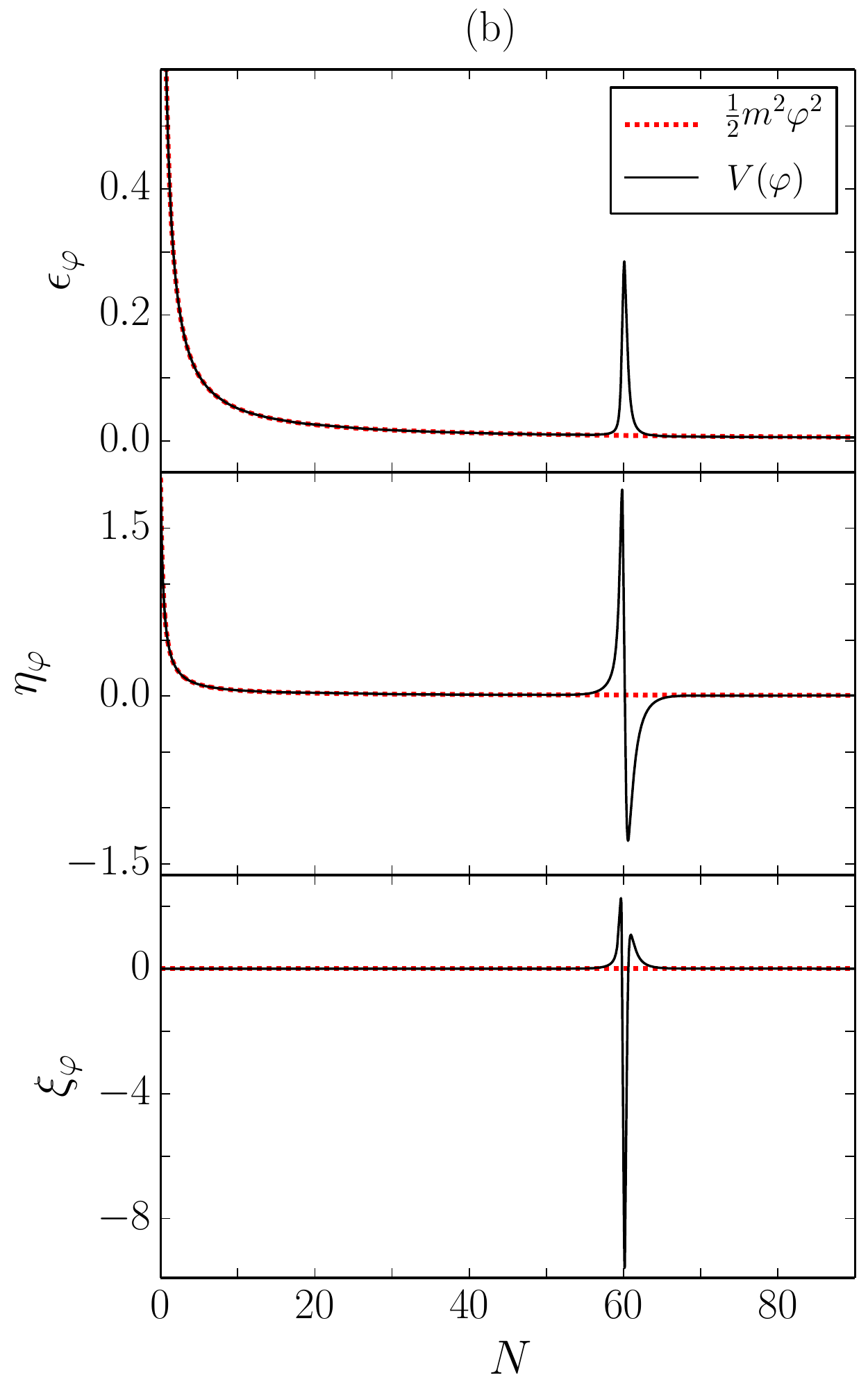}
\caption{Plots of the background parameters $H$, $\varphi$, and of the slow-roll
parameters $\epsilon$, $\epsilon_\varphi$, $\eta_\varphi$, and $\xi_\varphi$ in our model
(solid black curves) and in regular quadratic inflation (dotted red curves). The
horizontal axis represents the e-folding number $N$ with $N=0$ being set by the condition
$\epsilon(t_{\rm end})=1$, i.e.\ the end of inflation.}
\label{Fig:bg}
\end{center}
\end{figure*}

In this subsection, we perform the detailed analysis of the background inflationary
dynamics of the present model. It is well known that a model of $\varphi^2$ inflation
possesses a local attractor solution regardless of the initial condition for the
background scalar and that this is a slow-roll solution~\cite{Brandenberger:1990wu}.
Accordingly, in both the UV and IR regimes of our model, the inflaton field can enter a
period of slow-roll dynamics. Thus, in order to analyze the background dynamics
semi-analytically, it is convenient to introduce a series of slow-roll parameters as
follows,
\begin{align}
 & \epsilon \equiv -\frac{\dot{H}}{H^2} ~,~~
 \epsilon_\varphi \equiv \frac{M_p^2}{2} \Big( \frac{V_{,\varphi}}{V} \Big)^2 ~, \\
 & \eta_\varphi \equiv M_p^2 \frac{V_{,\varphi\varphi}}{V} ~,~~
 \xi_\varphi \equiv M_p^4 \frac{V_{,\varphi}V_{,\varphi\varphi\varphi}}{V^2} ~,
\end{align}
where a dot denotes a derivative with respect to cosmic time ($t$) and where the subscript
$_{,\varphi}$ denotes a derivative with respect to $\varphi$.

During inflation, the size of the universe is growing nearly exponentially while the
Hubble rate is slowly varying. Thus, instead of cosmic time, it is very efficient
to apply the inflationary e-folding number to characterize the time duration of the
process. Accordingly, the inflationary e-folding number is defined by
\begin{align}\label{Nefold}
 N(t) \equiv \int_t^{t_{\rm end}} H dt = \int^\varphi_{\varphi_{\rm end}} \frac{d\varphi}{\sqrt{2\epsilon}M_p}~,
\end{align}
where the end of inflation is determined by the condition $\epsilon(t_{\rm end}) = 1$.

When the single field $\varphi$ evolves stably along the slow-roll trajectory, it is not
difficult to observe that the value of $\epsilon$ approximately equals the value of
$\epsilon_\varphi$. One can easily derive the following approximate relations,
\begin{eqnarray}
 \epsilon_\varphi \simeq \eta_\varphi \simeq \frac{2M_p^2}{\varphi^2} ~, ~~ \xi_\varphi \simeq 0~,
\end{eqnarray}
when the inflaton field is far away from the critical scale.

When $\varphi$ evolves near the critical value $\varphi_c$, the potential
experiences a sudden decrease and its profile becomes very steep. In response to this
change in the potential, $\epsilon$ and $\epsilon_\varphi$ can increase significantly
before returning to their regular value after the critical phase. The slow-roll parameter
$|\eta_\varphi|$ can also obtain a dramatic increase around the critical scale but the
shape of the amplification is fairly asymmetric since this parameter characterizes the
second derivative of the potential. It is interesting to note that $|\xi_\varphi|$ in
standard quadratic inflation is always zero, but in our model, it can become very large
when the inflaton evolves through $\varphi_c$.

In the following, we numerically compute the background dynamics of our model.
In Fig.~\ref{Fig:bg}, we plot the evolution of the Hubble rate $H$, the inflaton field
$\varphi$, as well as the slow-roll parameters $\epsilon$, $\epsilon_\varphi$,
$\eta_\varphi$, and $\xi_\varphi$ with respect to the e-folding number $N$ (solid black
curves). Our model has four parameters, which are chosen to be, in Planck units,
\begin{eqnarray}
 &m_a=6.5175 \times 10^{-6}~,~~ \varphi_c=16~, \nonumber\\
 &m_b=7.9433 \times 10^{-6}~,~~  C_H=0.1~,
\end{eqnarray}
for this specific example. In order to make a comparison, we also compute the background
dynamics of quadratic inflation with a mass $m_a$ (dotted red curves). We note that we
set $N=0$ to be the end of inflation where $\epsilon(t_{\rm end})=1$. Thus, in these
figures, inflation begins on the right where all the slow-roll parameters are small
and proceeds to the left where both the Hubble rate and the inflaton field decrease to
smaller values.

From the plots in Fig.~\ref{Fig:bg}, one can clearly see that the behavior of the slow-roll
parameters is in agreement with the arguments given above. It is interesting to note that
when the inflaton field approaches $\varphi_c$, its evolution presents a small step
feature, and correspondingly, the Hubble rate experiences a sudden decrease at the same
moment. Even though this variation is very
small, it still leads to a series of changes in the slow-roll parameters. In particular,
$\xi_\varphi$, which characterizes the third derivative of the potential with respect to
$\varphi$, is vanishingly small for standard inflationary models. Here, we notice that
it can be amplified up to $\mathcal{O}(10)$.

We can see from Fig.~\ref{Fig:bg} that the deviations
between our model and the regular $m^2\varphi^2$ model are still very small.
Firstly, this is consistent with the nature of inflationary cosmology in which any
deviation from the attractor solution damps out quickly. Secondly, even though these
differences are very small, they can affect the process during which the primordial
perturbation modes exit the Hubble radius near the pivot scale.
We note that in our numerical computations, the pivot scale, which
is taken to be $k_* = 0.002~{\rm Mpc}^{-1}$, corresponds to the e-folding
number $N\simeq 55$. Accordingly, this can leave an imprint in the primordial power
spectra. This is the subject that we explore in the following subsection.

\subsection{Perturbation analysis}

\begin{figure}
\includegraphics[scale=0.5]{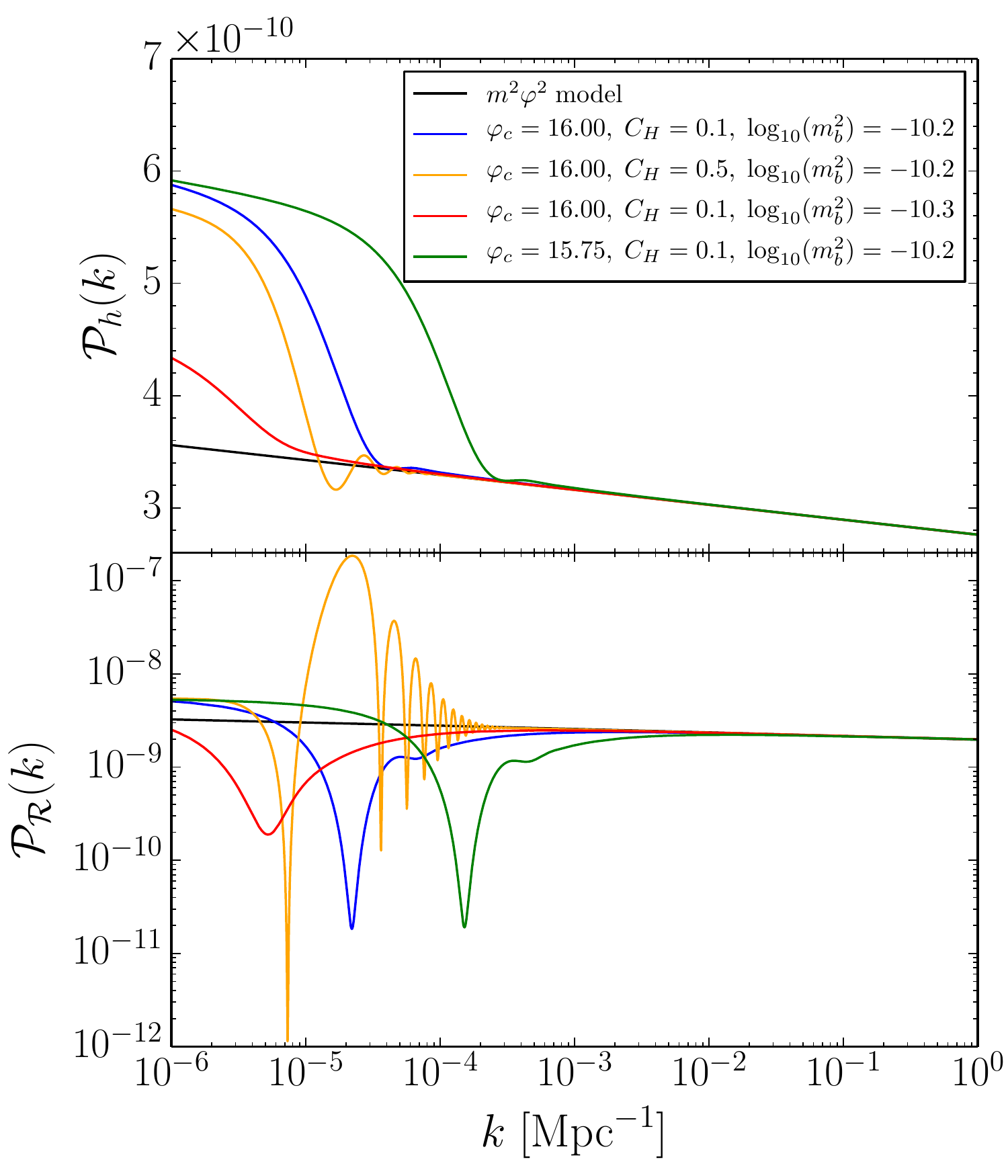}
\caption{Power spectrum for tensor modes ($\mathcal{P}_h$) and for curvature perturbations
($\mathcal{P}_{\mathcal{R}}$) as a function of the length scale of the perturbations in Fourier
space ($k$). The black curve shows the power spectra for standard quadratic inflation
where the mass of the inflaton is taken to be $m=m_a=6.5175\times 10^{-6}$. The colored
curves show the power spectra for the model presented in this section with different
parameter values.}
\label{Fig:powspec}
\end{figure}

We already see from the previous section that, at the background level, a step-like
feature in the potential induces a short period of time where the slow-roll conditions
are violated. We therefore expect this to have implications in the perturbative regime.
In fact, the near scale-invariant power spectra typically obtained from quadratic
inflation are likely to be modified, so in this section, we explore the quantitative
consequences of the model introduced above.

The power spectra for metric perturbations are usually found by solving the Einstein
field equations expanded about the background solution to linear order. For scalar modes,
a gauge-invariant perturbation quantity is the curvature perturbation $\mathcal{R}$, and
for tensor modes, it is the usual gravitational wave polarization state $h$ (see
\cite{Brandenberger:2003vk} for a review of cosmological perturbations). The respective
power spectra are then defined as
$\mathcal{P}_{\mathcal{R}}(k)\equiv k^3|\mathcal{R}_k|^2/(2\pi^2)$ and
$\mathcal{P}_h(k)\equiv k^3|h_k|^2/(2\pi^2)$. Given the complexity of the potential
energy for the scalar field and of the background dynamics, we solve the perturbations
equations for $\mathcal{R}$ and $h$ numerically in order to obtain their power spectra.
More specifically, we use the recently developed
\texttt{MultiModeCode}\footnote{\texttt{MultiModeCode} web page:\ \texttt{www.modecode.org}.}
\cite{Mortonson:2010er, Easther:2011yq, Price:2014xpa}. This code is optimized for
multi-field inflation, but its implementation makes it easy to evolve single field
inflation with the potential defined in Eq.~\eqref{V_para}.

\begin{figure*}
\includegraphics[scale=0.65]{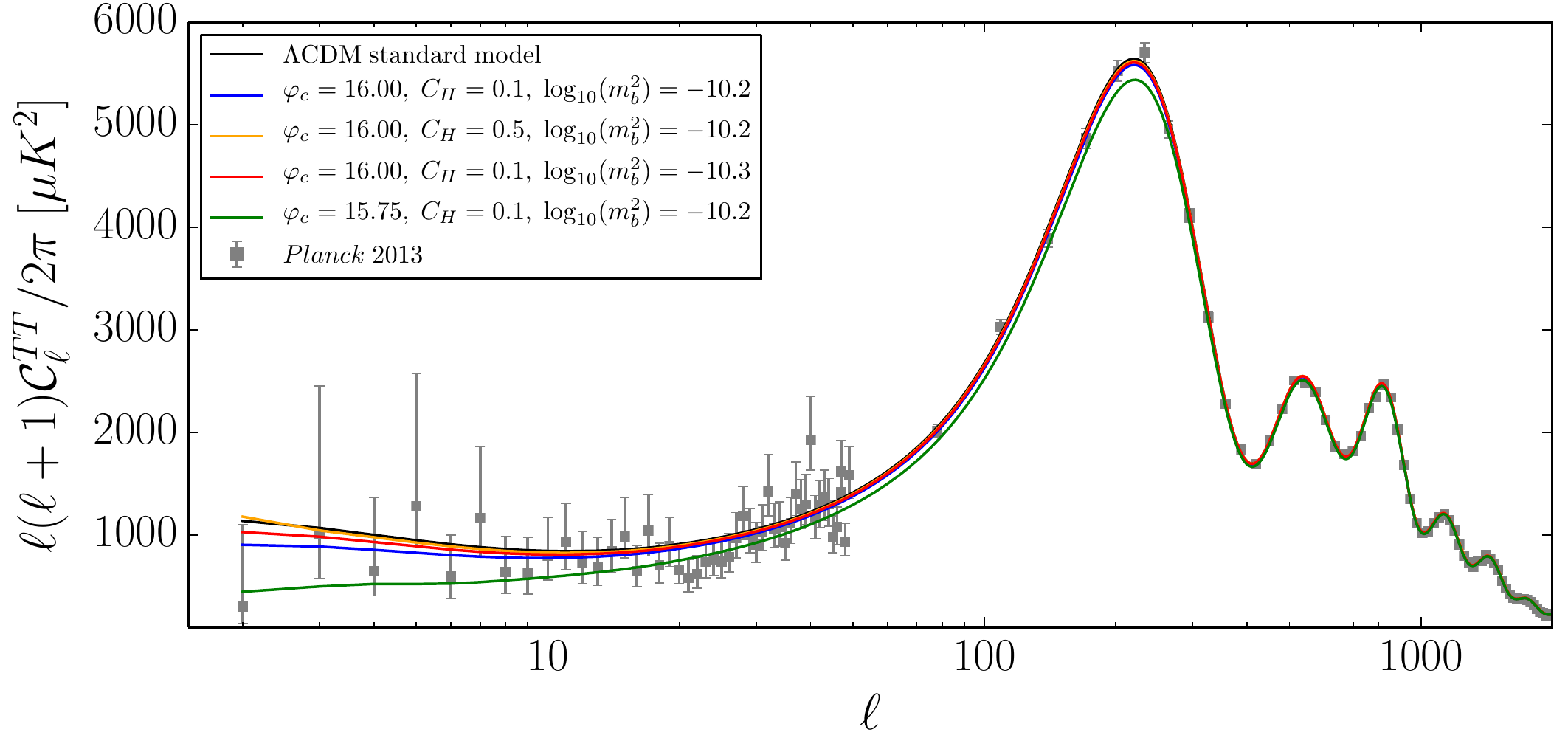}
\caption{Temperature (TT) angular power spectrum.
The grey squares show the results of the \textit{Planck} 2013 experiment and the black curve
shows the standard $\Lambda$CDM model. The colored curves show the angular power spectrum for
the model presented in this section with different parameter values. The value
$m_a=6.5175\times 10^{-6}$ is fixed. The value of the cosmological parameters (e.g.,
$\Omega_bh^2$, $\Omega_ch^2$, $H_0$) are taken from the \textit{Planck} 2013 collaboration and
they are fixed at the same value for every curve. Furthermore, for the $\Lambda$CDM curve,
the primordial power spectrum is parameterized by
$\mathcal{P}_{\mathcal{R}}(k)=A_s(k/k_*)^{n_s-1}$ where we take
$A_s=2.21536\times 10^{-9}$, $k_*=0.05~\mathrm{Mpc}^{-1}$, and $n_s=0.96$.}
\label{Fig:angpowspec}
\end{figure*}

The resulting power spectra are shown in Fig.~\ref{Fig:powspec}. We plotted the standard
near scale-invariant power spectra for quadratic inflation in black, and the color curves
show the modified potential with different parameter values. We first
notice that for $k\gtrsim 10^{-2}~\mathrm{Mpc}^{-1}$, the standard $m^2\varphi^2$
potential and the modified potential yield near identical power spectra. Moreover,
for the parameter values shown in Fig.~\ref{Fig:powspec}, our model predicts a similar
tilt and tensor-to-scalar ratio compared to quadratic inflation with $n_s\approx 0.96$
and $r\approx 0.14$ at the pivot scale of $k_*=0.05~\mathrm{Mpc}^{-1}$. It is only on
larger length scales that the spectra deviate from near scale invariance due to the short phase
of slow-roll violation during the inflationary period. Overall, the modified potential
tends to lower the amplitude of the curvature perturbations and enhance the amount of
gravitational waves produced. The effect is more radical when the drop in the potential
depicted in Fig.~\ref{Fig:sketch} is very steep. In our parametrization of the potential,
this is equivalent to having a large value of $C_H$. This is shown by the orange curve
which has small damped non-harmonic oscillations for tensor modes but very large ones in
the curvature power spectrum. However, such a large value for $C_H$ is not expected from
string theory (see Sec.~\ref{sec:newaxion}).

The effect is less drastic when the drop
in the potential is smooth and this can be seen with the curves in red, blue, and green.
In particular, the smallest effect is found when the large mass scale $m_b$ is close to
$m_a$ (red curve). In this case, the jump in the potential is small and the feature in
the power spectra occurs only in the far infrared. Finally, we note that changing the
critical field value $\varphi_c$ at which the jump in the potential occurs affects the
scale at which the power spectra leave near scale-invariance. More specifically, the
lower the energy scale of the drop in the potential, the lower the length scale of the feature in the
power spectra.

Let us provide some further physical intuition for the above description of the
power spectra. The small $k$ modes (large wavelength) exit the Hubble radius at
early times when the inflaton is in its first slow-roll phase, i.e.\ when the potential
looks like $m_b^2\varphi^2/2$. At this point, the power spectra are given by their usual
forms~\cite{Mukhanov:1990me} with an amplitude proportional to
$H^2/(\epsilon M_p^2)$ for curvature perturbations and proportional to $H^2/M_p^2$
for tensor modes. Thus, the amplitudes depend on $m_b^2$ (the larger mass scale).
Similarly, the large $k$ modes (small wavelength) exit the Hubble radius at late
times when the inflaton is in its second slow-roll phase, i.e.\ when the potential
looks like $m_a^2\varphi^2/2$, so the amplitude of the spectra depends on $m_a^2$
(the smaller mass scale). In particular, the amplitude of the spectra at the pivot
scale $k_*=0.05~\mathrm{Mpc}^{-1}$ is most sensitive to the mass parameter $m_a$.
Finally, modes of intermediate wavelengths exit the Hubble radius near the critical scale
$\varphi_c$ at which point the slow-roll conditions are violated. Consequently, the power
spectra do not follow their usual near scale-invariance for slow-roll inflation.
The behavior of the power spectra in this regime depends on the parameter values of the
model and has been described for a number of cases in the previous paragraph.

To further connect with observations, we evolve the resulting power spectra after
inflation to the angular power spectra that can be probed observationally at the time of
the CMB with telescopes such as \textit{Planck}. To do this, we
use \texttt{CAMB}\footnote{\texttt{CAMB} web page:\ \texttt{http://camb.info}.}
\cite{Lewis:1999bs} where we input our power spectrum for curvature perturbations and
tensor modes as shown in Fig.~\ref{Fig:powspec} instead of a parameterized power spectrum
as it is usually done for the standard $\Lambda$CDM model. Furthermore, we use the best
fit to the cosmological parameters found by\ \textit{Planck}~\cite{Ade:2013zuv} in the
code.

First, the temperature (TT) angular power spectrum is shown in Fig.~\ref{Fig:angpowspec}.
The grey squares show the measured data by \textit{Planck} 2013 and the black curve shows the
standard $\Lambda$CDM curve for a parameterized power spectrum after inflation. The color curves
show the angular power spectrum for the modified potential, and similar to our conclusion
from the power spectra after inflation, the different angular power spectra are very
similar on small scales ($\ell\gtrsim 300$).

The differences with $\Lambda$CDM are more
distinct around the first acoustic peak and at low $\ell$'s. In particular, we notice
that the amplitude of the green curve is much smaller than the other ones for
$\ell\lesssim 300$. This was expected from the fact that the curvature power spectrum for
the green curve is the one that deviates from near scale-invariance at the smallest scale
(see lower panel of Fig.~\ref{Fig:powspec}) since it has the smallest $\varphi_c$.
Although the low amplitude of the angular power spectrum around the
first acoustic peak may imply larger differences in the cosmological parameters compared
to the \textit{Planck} 2013 results, the green curve has the advantage of explaining the
low-$\ell$ suppression of the spectrum. Such a suppression is also obtained in the case of
the blue and red curves, although at a smaller extent than for the case of the green curve,
but these curves have the advantage of closely matching $\Lambda$CDM around the first
acoustic peak. To summarize, the larger the $k$-mode at which the feature appears in the
curvature power spectrum, the larger the deviation from $\Lambda$CDM, but the larger the
low-$\ell$ suppression is.

The situation is somewhat different for
the orange curve in the case of a steep drop in the potential (large $C_H$). Indeed,
this case follows $\Lambda$CDM very closely at all angular scales except at $\ell=2$ where
the amplitude for the modified potential is larger and this seems to be in greater
conflict with the observation. This reinforces our intuition that a steep drop in the
potential is not favored.

\begin{figure}
\includegraphics[scale=0.40]{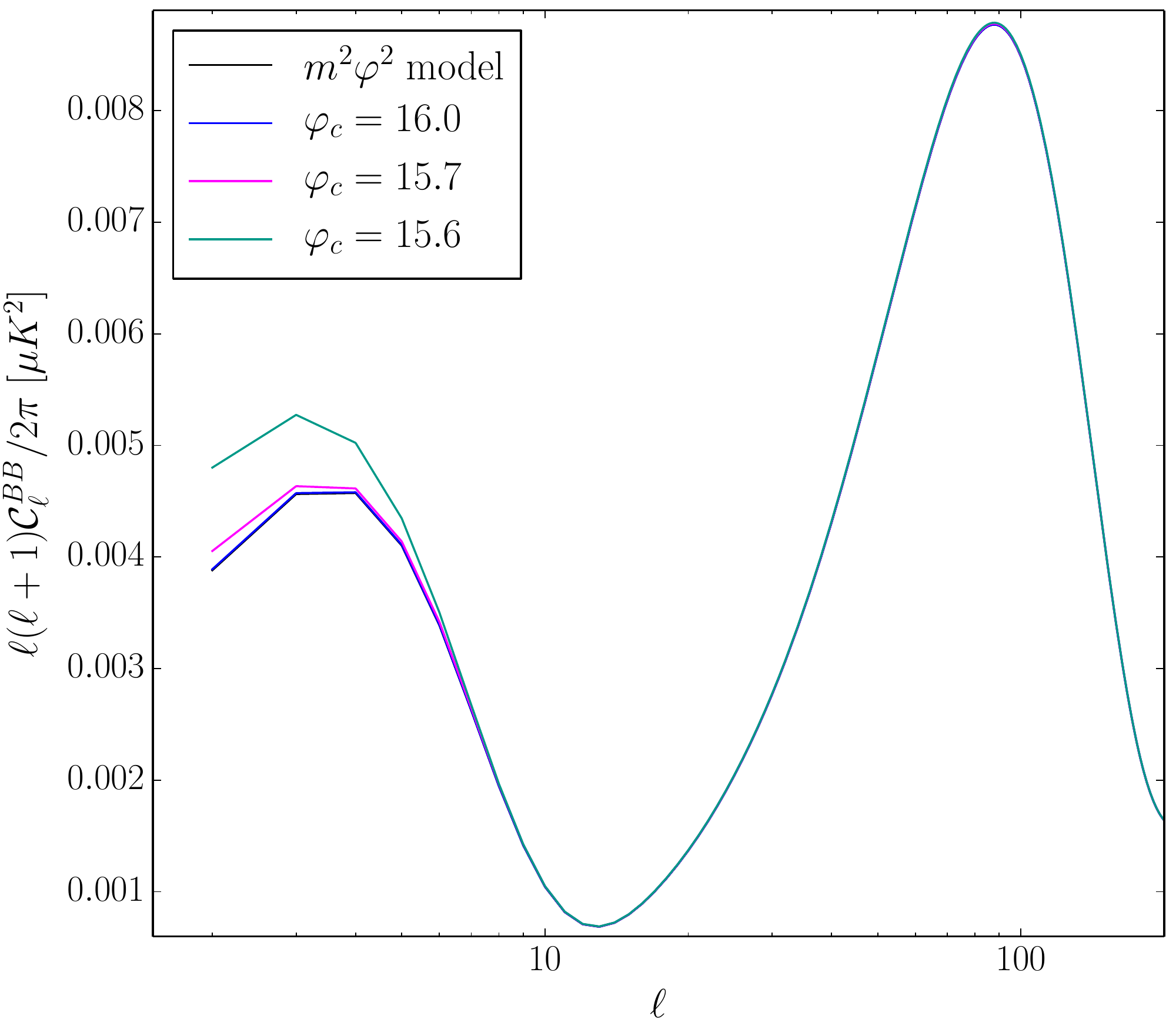}
\caption{Unlensed B-mode polarization angular power spectrum.
The black curve shows the standard result for quadratic inflation.
The blue, magenta, and turquoise curves show the result for our model
when the critical scale $\varphi_c$ changes. The other parameters are
kept fixed: $m_a=6.5175 \times 10^{-6}$, $m_b=7.9433 \times 10^{-6}$, $C_H=0.1$.
We note that the black curve is completely hidden below the blue curve.}
\label{Fig:angpowspecBB}
\end{figure}

\begin{figure*}
\includegraphics[scale=0.45]{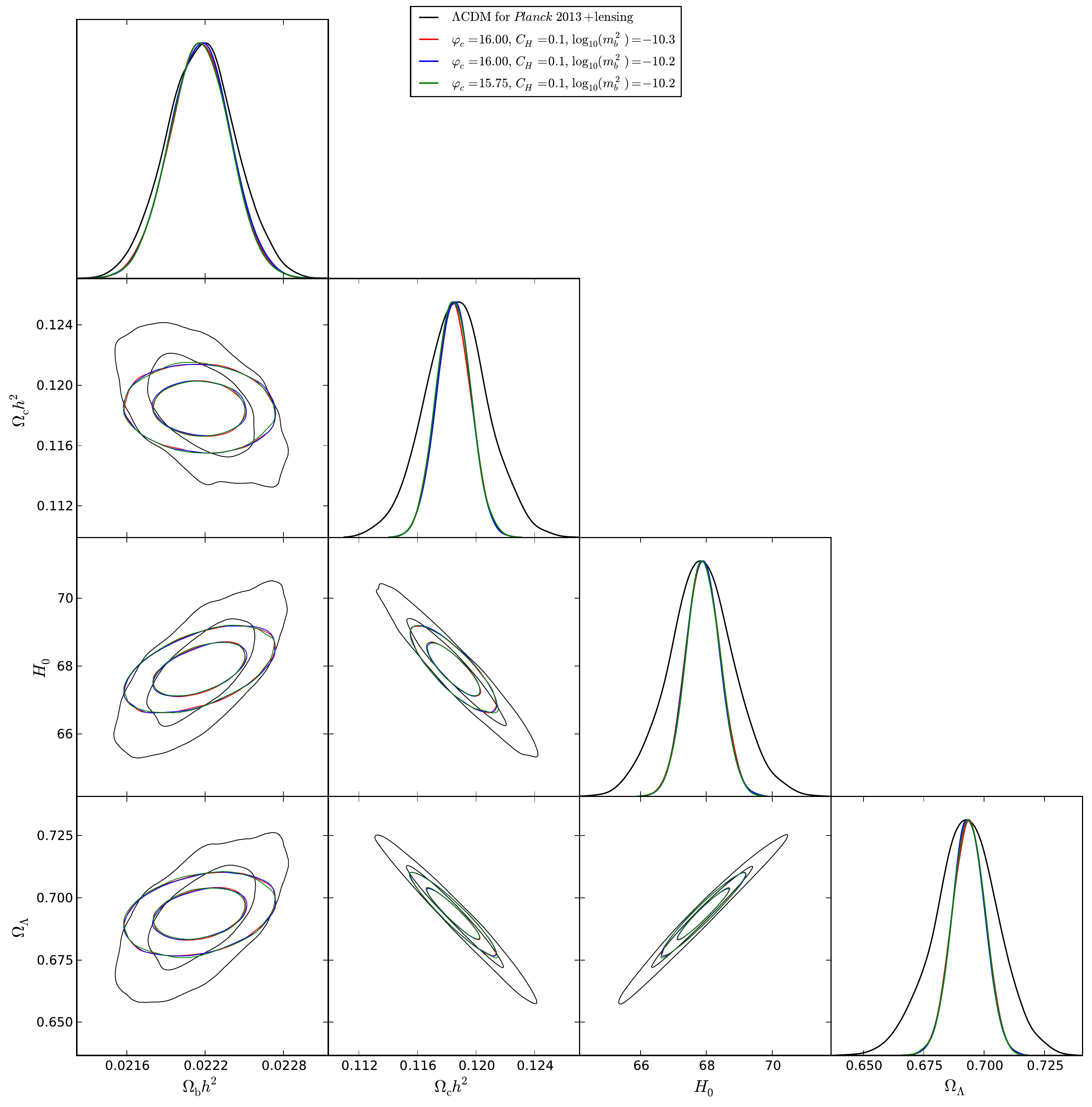}
\caption{Constraints on pairs of cosmological parameters (1$\sigma$ and 2$\sigma$
contour lines) and probability distributions for the baryon density today
($\Omega_bh^2$), the cold dark matter density today ($\Omega_ch^2$), the Hubble parameter
today ($H_0$), and the dark energy density divided by the critical density today
($\Omega_{\Lambda}$). The color scheme is the same as for Fig.~\ref{Fig:angpowspec}.}
\label{Fig:CosmoMC}
\end{figure*}

Second, we show the unlensed B-mode polarization angular power spectrum in
Fig.~\ref{Fig:angpowspecBB}, i.e.\ the primordial contribution to the total B-mode power
spectrum. The figure shows that the axion monodromy model with different parameter values
agrees with the quadratic potential at almost all angular scales. In fact, for $\ell\gtrsim 7$,
the spectra are basically identical. This is in agreement with the top panel of
Fig.~\ref{Fig:powspec} which showed very similar tensor power spectra for large $k$.

It is only at very large angular scales ($\ell\lesssim 7$) that one can notice a small
deviation. This deviation is only significant when the critical scale $\varphi_c$
is small since this is when the deviation from near scale invariance in the primordial
power spectrum occurs at larger $k$. We notice that the amplification at small $\ell$ in
the BB power spectrum is very small compared to the suppression in the TT power spectrum.
This is in accordance with the power spectra after inflation shown in
Fig.~\ref{Fig:powspec} where the suppression in $\mathcal{P}_{\mathcal{R}}$ can be of more
than one order of magnitude, whereas the amplification in $\mathcal{P}_h$ is only of order
one. Clearly, the suppression in the TT spectrum and the amplification in the BB spectrum
are correlated, so one needs to compare both spectra with observations to constrain the
model. Still, the amplification at large angular scales in the BB spectrum is an
interesting feature of this model and one hopes to explore this region as well as the
entire spectrum with future CMB polarization data.

\subsection{Fitting to cosmological data}

After analyzing the temperature angular power spectrum, one would hope to be able to
determine which model best fits the observed data. In Fig.~\ref{Fig:angpowspec}, the
cosmological parameters are fixed, but when one allows them to be free, it is possible
to fit all the cosmological parameters to the data given the model that is studied. This
is what we explore with Markov Chain Monte Carlo (MCMC) simulations.

We perform a model comparison with
\texttt{CosmoMC}\footnote{\texttt{CosmoMC} web page:\ \texttt{http://cosmologist.info/cosmomc}.}
\cite{Lewis:2002ah} by running the MCMC code for the red, blue, and green curves
of Fig.~\ref{Fig:angpowspec}. The usual cosmological parameters (except the ones related
to a parameterized power spectrum such as $n_s$ and $A_s$, which are ignored) are given
the same prior range as \textit{Planck}~\cite{Ade:2013zuv}. The runs reached convergency
as we can see from the Gelman-Rubin statistic $R-1$ value
\cite{Gelman:1992zz,Lewis:2013hha} of 0.01325, 0.01276, and 0.01103 in the case of the
red, blue, and green curves, respectively. The resulting likelihoods can then be compared
to determine which of the three cases best fits the \textit{Planck} 2013 data, but we find
that the likelihoods are all identical within the precision of the results. Furthermore,
running the action 2 from \texttt{CosmoMC} leads to the same conclusion: the respective
$\chi^2$ are near identical and it is therefore not possible to tell which case is best.

For this reason, we leave model comparison aside and simply compare the\ \texttt{CosmoMC}
results for our model with the standard $\Lambda$CDM model from \textit{Planck} 2013. In
Fig.~\ref{Fig:CosmoMC}, we show constraints on pairs of cosmological parameters and their
respective probability distributions for the\ \texttt{CosmoMC} runs described above and
for the results from \textit{Planck} 2013. We first notice that the colored curve which
represent our model are almost perfectly superposed, in agreement with our likelihood
analysis. Also, we notice that the resulting best fits to the cosmological parameters are
all very close to one another even when comparing with the \textit{Planck} 2013 results. They
differ by fractions of the order of 1\% at most. This suggests that the model tested here
is in good agreement with the observations. We stress though that the above analysis
cannot indicate which of the axion monodromy model and the standard model is best.
Indeed, the standard model assumes a parameterized power spectrum of the form
$\mathcal{P}_{\mathcal{R}}(k)=A_s(k/k_*)^{n_s-1}$ and thus fits two additional
parameters, i.e.\ $A_s$ and $n_s$. This in part explains why the cosmological parameters
in Fig.~\ref{Fig:CosmoMC} are better constrained by our model than by the standard model
and why more correlation seems present (especially for $\Omega_bh^2$) in the standard
model than in the axion monodromy model.

\section{Conclusions}\label{sec:conclusion}

In the present paper, we proposed a new model of axion monodromy inflation.
Analogously to regular models of axion monodromy inflation in which the
effective theory is derived from the reduction of a higher dimensional
theory on a two-cycle, our model is constructed in the presence of
torsional cycles. In particular, our model possesses at least three
scalar fields that can be candidates for the inflaton. Amongst them,
two fields are free of the $\eta$ problem due to mass corrections of
the moduli stabilization in the K\"{a}hler potential, and hence,
the theory is well established from the perspective of string theory.

As a particular realization of this model, we studied quadratic inflation
where the potential is effectively transformed
to have two mass scales separated by a step-like feature at some critical
scale. At the background level, one finds a short phase in which the
slow-roll conditions are violated. Perturbatively, the power spectrum of
curvature perturbations is either suppressed or oscillates near the
critical scale depending on the details of the potential, and the power spectrum
of tensor modes is boosted on large scales.

In light of recent and forthcoming CMB experiments, a lot of analyses have aimed at searching
for features of inflation or its alternatives (see, e.g.,
\cite{Contaldi:2014zua,Ashoorioon:2014nta,Feng:2003mk,Wan:2014fra,Miranda:2014wga,Firouzjahi:2014fda,Piao:2003zm,Xia:2014tda,Cai:2014hja,Cai:2014bea,Joy:2007na,Joy:2008qd,Hamann:2007pa,Jain:2008dw,Jain:2009pm,Mortonson:2009qv,Hazra:2010ve,Hazra:2014jka,Hazra:2014goa,Hazra:2013nca,Abazajian:2014tqa,Hu:2014aua,Ashoorioon:2006wc,Ashoorioon:2008qr,Ashoorioon:2014yua,Ballesteros:2014yva,Cai:2015dta}
for extensive discussions). In our model, we have two interesting predictions:
one is that there exits a moderate boost in the CMB B-mode polarization spectrum at small $\ell$'s;
the other is a defect of temperature anisotropies in the same regime,
which provides a natural interpretation for the low-$\ell$ suppression anomaly in the CMB data.
It will be interesting to perform a full global fitting analysis with the\ \textit{Planck} 2015 data
to confront these predictions with cosmological observations, which will be presented in
a follow-up letter~\cite{Cai:2015xla}.

\vspace{0.5cm}

We point out that while this paper was being prepared for submission,
the preprint of~\cite{Flauger:2014ana} appeared, which explores features
of standard axion monodromy inflation with drifting oscillations.

\section*{Acknowledgments}

We thank Robert Brandenberger, Jim Cline, Gil Holder, Bin Hu, Gary Shiu, Gabrielle Simard, Anzhong Wang,
and in particular, Keshav Dasgupta, Evan McDonough, and Gim Seng Ng for valuable discussions.
YFC is supported in part by the Natural Sciences and Engineering Research Council (NSERC) of Canada,
by the Department of Physics at McGill University, by the Chinese National Youth Thousand Talents Program,
by the USTC start-up funding (KY2030000049), and by the Natural Science Foundation of China (Grant No. 11421303).
FC is supported by National Science Foundation (NSF) Grant No. PHY11-25915.
EF thanks CNPq (Science without Borders) for financial support.
JQ acknowledges the support, throughout the completion of this work,
of the Fonds de recherche du Qu\'{e}bec - Nature et technologies (FRQNT),
the Walter C. Sumner Foundation,
and NSERC via the Vanier Canada Graduate Scholarships program.
Computations were made on the supercomputer Guillimin from McGill University,
managed by Calcul Qu\'{e}bec and Compute Canada. The operation of this supercomputer is
funded by the Canada Foundation for Innovation (CFI), NanoQu\'{e}bec, RMGA and FRQNT.


\begin{thebibliography}{999}

\bibitem{Ade:2013zuv}
  P.~A.~R.~Ade {\it et al.}
  [Planck Collaboration],
  ``Planck 2013 results. XVI. Cosmological parameters,''
  Astron.\ Astrophys.\  {\bf 571}, A16 (2014)
  [arXiv:1303.5076 [astro-ph.CO]].

\bibitem{Planck:2013jfk}
  P.~A.~R.~Ade {\it et al.} [Planck Collaboration],
  ``Planck 2013 results. XXII. Constraints on inflation,''
  Astron.\ Astrophys.\  {\bf 571}, A22 (2014)
  [arXiv:1303.5082 [astro-ph.CO]].

\bibitem{Ade:2014xna}
  P.~A.~R.~Ade {\it et al.}
  [BICEP2 Collaboration],
  ``Detection of B-Mode Polarization at Degree Angular Scales by BICEP2,''
  Phys.\ Rev.\ Lett.\  {\bf 112}, 241101 (2014)
  [arXiv:1403.3985 [astro-ph.CO]].

\bibitem{Lyth:1996im}
  D.~H.~Lyth,
  ``What would we learn by detecting a gravitational wave signal in the cosmic microwave background anisotropy?,''
  Phys.\ Rev.\ Lett.\  {\bf 78}, 1861 (1997)
  [hep-ph/9606387].

\bibitem{Mortonson:2014bja}
  M.~J.~Mortonson and U.~Seljak,
  ``A joint analysis of Planck and BICEP2 B modes including dust polarization uncertainty,''
  JCAP {\bf 1410}, no. 10, 035 (2014)
  [arXiv:1405.5857 [astro-ph.CO]].

\bibitem{Flauger:2014qra}
  R.~Flauger, J.~C.~Hill and D.~N.~Spergel,
  ``Toward an Understanding of Foreground Emission in the BICEP2 Region,''
  JCAP {\bf 1408}, 039 (2014)
  [arXiv:1405.7351 [astro-ph.CO]].

\bibitem{Adam:2014bub}
  R.~Adam {\it et al.} [Planck Collaboration],
  ``Planck intermediate results. XXX. The angular power spectrum of polarized dust emission at intermediate and high Galactic latitudes,''
  Astron.\ Astrophys.\  {\bf 586}, A133 (2016)
  [arXiv:1409.5738 [astro-ph.CO]].

\bibitem{Burgess:2013sla}
  C.~P.~Burgess, M.~Cicoli and F.~Quevedo,
  ``String Inflation After Planck 2013,''
  JCAP {\bf 1311}, 003 (2013)
  [arXiv:1306.3512 [hep-th]].

\bibitem{Baumann:2014nda}
  D.~Baumann and L.~McAllister,
  ``Inflation and String Theory,''
  arXiv:1404.2601 [hep-th].

\bibitem{Westphal:2014ana}
  A.~Westphal,
  ``String cosmology \textemdash\ Large-field inflation in string theory,''
  Int.\ J.\ Mod.\ Phys.\ A {\bf 30}, no. 09, 1530024 (2015)
  [arXiv:1409.5350 [hep-th]].

\bibitem{Silverstein:2008sg}
  E.~Silverstein and A.~Westphal,
  ``Monodromy in the CMB: Gravity Waves and String Inflation,''
  Phys.\ Rev.\ D {\bf 78}, 106003 (2008)
  [arXiv:0803.3085 [hep-th]].

\bibitem{McAllister:2008hb}
  L.~McAllister, E.~Silverstein and A.~Westphal,
  ``Gravity Waves and Linear Inflation from Axion Monodromy,''
  Phys.\ Rev.\ D {\bf 82}, 046003 (2010)
  [arXiv:0808.0706 [hep-th]].

\bibitem{Flauger:2009ab}
  R.~Flauger, L.~McAllister, E.~Pajer, A.~Westphal and G.~Xu,
  ``Oscillations in the CMB from Axion Monodromy Inflation,''
  JCAP {\bf 1006}, 009 (2010)
  [arXiv:0907.2916 [hep-th]].

\bibitem{McAllister:2014mpa}
  L.~McAllister, E.~Silverstein, A.~Westphal and T.~Wrase,
  ``The Powers of Monodromy,''
  JHEP {\bf 1409}, 123 (2014)
  [arXiv:1405.3652 [hep-th]].

\bibitem{Marchesano:2014mla}
  F.~Marchesano, G.~Shiu and A.~M.~Uranga,
  ``F-term Axion Monodromy Inflation,''
  JHEP {\bf 1409}, 184 (2014)
  [arXiv:1404.3040 [hep-th]].

\bibitem{Blumenhagen:2014gta}
  R.~Blumenhagen and E.~Plauschinn,
  ``Towards Universal Axion Inflation and Reheating in String Theory,''
  Phys.\ Lett.\ B {\bf 736}, 482 (2014)
  [arXiv:1404.3542 [hep-th]].

\bibitem{Hebecker:2014eua}
  A.~Hebecker, S.~C.~Kraus and L.~T.~Witkowski,
  ``D7-Brane Chaotic Inflation,''
  Phys.\ Lett.\ B {\bf 737}, 16 (2014)
  [arXiv:1404.3711 [hep-th]].

\bibitem{Ibanez:2014kia}
  L.~E.~Ib\'{a}\~{n}ez and I.~Valenzuela,
  ``The inflaton as an MSSM Higgs and open string modulus monodromy inflation,''
  Phys.\ Lett.\ B {\bf 736}, 226 (2014)
  [arXiv:1404.5235 [hep-th]].

\bibitem{Franco:2014hsa}
  S.~Franco, D.~Galloni, A.~Retolaza and A.~Uranga,
  ``On axion monodromy inflation in warped throats,''
  JHEP {\bf 1502}, 086 (2015)
  [arXiv:1405.7044 [hep-th]].

\bibitem{Freese:1990rb}
  K.~Freese, J.~A.~Frieman and A.~V.~Olinto,
  ``Natural inflation with pseudo - Nambu-Goldstone bosons,''
  Phys.\ Rev.\ Lett.\  {\bf 65}, 3233 (1990).

\bibitem{Kim:2004rp}
  J.~E.~Kim, H.~P.~Nilles and M.~Peloso,
  ``Completing natural inflation,''
  JCAP {\bf 0501}, 005 (2005)
  [hep-ph/0409138].

\bibitem{Kappl:2014lra}
  R.~Kappl, S.~Krippendorf and H.~P.~Nilles,
  ``Aligned Natural Inflation: Monodromies of two Axions,''
  Phys.\ Lett.\ B {\bf 737}, 124 (2014)
  [arXiv:1404.7127 [hep-th]].

\bibitem{Long:2014dta}
  C.~Long, L.~McAllister and P.~McGuirk,
  ``Aligned Natural Inflation in String Theory,''
  Phys.\ Rev.\ D {\bf 90}, 023501 (2014)
  [arXiv:1404.7852 [hep-th]].

\bibitem{Bachlechner:2014hsa}
  T.~C.~Bachlechner, M.~Dias, J.~Frazer and L.~McAllister,
  ``Chaotic inflation with kinetic alignment of axion fields,''
  Phys.\ Rev.\ D {\bf 91}, no. 2, 023520 (2015)
  [arXiv:1404.7496 [hep-th]].

\bibitem{Blumenhagen:2014nba}
  R.~Blumenhagen, D.~Herschmann and E.~Plauschinn,
  ``The Challenge of Realizing F-term Axion Monodromy Inflation in String Theory,''
  JHEP {\bf 1501}, 007 (2015)
  [arXiv:1409.7075 [hep-th]].

\bibitem{Hayashi:2014aua}
  H.~Hayashi, R.~Matsuda and T.~Watari,
  ``Issues in Complex Structure Moduli Inflation,''
  arXiv:1410.7522 [hep-th].

\bibitem{Hebecker:2014kva}
  A.~Hebecker, P.~Mangat, F.~Rompineve and L.~T.~Witkowski,
  ``Tuning and Backreaction in F-term Axion Monodromy Inflation,''
  Nucl.\ Phys.\ B {\bf 894}, 456 (2015)
  [arXiv:1411.2032 [hep-th]].

\bibitem{Berg:2009tg}
  M.~Berg, E.~Pajer and S.~Sjors,
  ``Dante's Inferno,''
  Phys.\ Rev.\ D {\bf 81}, 103535 (2010)
  [arXiv:0912.1341 [hep-th]].

\bibitem{Hannestad:2009yx}
  S.~Hannestad, T.~Haugbolle, P.~R.~Jarnhus and M.~S.~Sloth,
  ``Non-Gaussianity from Axion Monodromy Inflation,''
  JCAP {\bf 1006}, 001 (2010)
  [arXiv:0912.3527 [hep-ph]].

\bibitem{Conlon:2011qp}
  J.~P.~Conlon,
  ``Brane-Antibrane Backreaction in Axion Monodromy Inflation,''
  JCAP {\bf 1201}, 033 (2012)
  [arXiv:1110.6454 [hep-th]].

\bibitem{Shlaer:2012by}
  B.~Shlaer,
  ``Chaotic Brane Inflation,''
  Phys.\ Rev.\ D {\bf 88}, 103503 (2013)
  [arXiv:1211.4024 [hep-th]].

\bibitem{Gao:2014uha}
  X.~Gao, T.~Li and P.~Shukla,
  ``Combining Universal and Odd RR Axions for Aligned Natural Inflation,''
  JCAP {\bf 1410}, no. 10, 048 (2014)
  [arXiv:1406.0341 [hep-th]].

\bibitem{Ozsoy:2014sba}
  O.~\"{O}zsoy, K.~Sinha and S.~Watson,
  ``How Well Can We Really Determine the Scale of Inflation?,''
  Phys.\ Rev.\ D {\bf 91}, no. 10, 103509 (2015)
  [arXiv:1410.0016 [hep-th]].

\bibitem{Kaloper:2008fb}
  N.~Kaloper and L.~Sorbo,
  ``A Natural Framework for Chaotic Inflation,''
  Phys.\ Rev.\ Lett.\  {\bf 102}, 121301 (2009)
  [arXiv:0811.1989 [hep-th]].

\bibitem{Copeland:1994vg}
  E.~J.~Copeland, A.~R.~Liddle, D.~H.~Lyth, E.~D.~Stewart and D.~Wands,
  ``False vacuum inflation with Einstein gravity,''
  Phys.\ Rev.\ D {\bf 49}, 6410 (1994)
  [astro-ph/9401011].

\bibitem{DeWolfe:2002nn}
  O.~DeWolfe and S.~B.~Giddings,
  ``Scales and hierarchies in warped compactifications and brane worlds,''
  Phys.\ Rev.\ D {\bf 67}, 066008 (2003)
  [hep-th/0208123].

\bibitem{deAlwis:2003sn}
  S.~P.~de Alwis,
  ``On Potentials from fluxes,''
  Phys.\ Rev.\ D {\bf 68}, 126001 (2003)
  [hep-th/0307084].

\bibitem{Giddings:2005ff}
  S.~B.~Giddings and A.~Maharana,
  ``Dynamics of warped compactifications and the shape of the warped landscape,''
  Phys.\ Rev.\ D {\bf 73}, 126003 (2006)
  [hep-th/0507158].

\bibitem{Shiu:2008ry}
  G.~Shiu, G.~Torroba, B.~Underwood and M.~R.~Douglas,
  ``Dynamics of Warped Flux Compactifications,''
  JHEP {\bf 0806}, 024 (2008)
  [arXiv:0803.3068 [hep-th]].

\bibitem{Douglas:2008jx}
  M.~R.~Douglas and G.~Torroba,
  ``Kinetic terms in warped compactifications,''
  JHEP {\bf 0905}, 013 (2009)
  [arXiv:0805.3700 [hep-th]].

\bibitem{Frey:2008xw}
  A.~R.~Frey, G.~Torroba, B.~Underwood and M.~R.~Douglas,
  ``The Universal Kahler Modulus in Warped Compactifications,''
  JHEP {\bf 0901}, 036 (2009)
  [arXiv:0810.5768 [hep-th]].

\bibitem{Underwood:2010pm}
  B.~Underwood,
  ``A Breathing Mode for Warped Compactifications,''
  Class.\ Quant.\ Grav.\  {\bf 28}, 195013 (2011)
  [arXiv:1009.4200 [hep-th]].

\bibitem{Anguelova:2010ed}
  L.~Anguelova, C.~Quigley and S.~Sethi,
  ``The Leading Quantum Corrections to Stringy Kahler Potentials,''
  JHEP {\bf 1010}, 065 (2010)
  [arXiv:1007.4793 [hep-th]].

\bibitem{Kachru:2003aw}
  S.~Kachru, R.~Kallosh, A.~D.~Linde and S.~P.~Trivedi,
  ``De Sitter vacua in string theory,''
  Phys.\ Rev.\ D {\bf 68}, 046005 (2003)
  [hep-th/0301240].

\bibitem{Giddings:2001yu}
  S.~B.~Giddings, S.~Kachru and J.~Polchinski,
  ``Hierarchies from fluxes in string compactifications,''
  Phys.\ Rev.\ D {\bf 66}, 106006 (2002)
  [hep-th/0105097].

\bibitem{Dasgupta:2014pma}
  K.~Dasgupta, R.~Gwyn, E.~McDonough, M.~Mia and R.~Tatar,
  ``de Sitter Vacua in Type IIB String Theory: Classical Solutions and Quantum Corrections,''
  JHEP {\bf 1407}, 054 (2014)
  [arXiv:1402.5112 [hep-th]].

\bibitem{Douglas:2006es}
  M.~R.~Douglas and S.~Kachru,
  ``Flux compactification,''
  Rev.\ Mod.\ Phys.\  {\bf 79}, 733 (2007)
  [hep-th/0610102].

\bibitem{Denef:2007pq}
  F.~Denef, M.~R.~Douglas and S.~Kachru,
  ``Physics of String Flux Compactifications,''
  Ann.\ Rev.\ Nucl.\ Part.\ Sci.\  {\bf 57}, 119 (2007)
  [hep-th/0701050].

\bibitem{Starobinsky:1992ts}
  A.~A.~Starobinsky,
  ``Spectrum of adiabatic perturbations in the universe when there are singularities in the inflation potential,''
  JETP Lett.\  {\bf 55}, 489 (1992)
  [Pisma Zh.\ Eksp.\ Teor.\ Fiz.\  {\bf 55}, 477 (1992)].

\bibitem{Adams:1997de}
  J.~A.~Adams, G.~G.~Ross and S.~Sarkar,
  ``Multiple inflation,''
  Nucl.\ Phys.\ B {\bf 503}, 405 (1997)
  [hep-ph/9704286].

\bibitem{Adams:2001vc}
  J.~A.~Adams, B.~Cresswell and R.~Easther,
  ``Inflationary perturbations from a potential with a step,''
  Phys.\ Rev.\ D {\bf 64}, 123514 (2001)
  [astro-ph/0102236].

\bibitem{Feng:2003zua}
  B.~Feng and X.~Zhang,
  ``Double inflation and the low cmb quadrupole,''
  Phys.\ Lett.\ B {\bf 570}, 145 (2003)
  [astro-ph/0305020].

\bibitem{Joy:2007na}
  M.~Joy, V.~Sahni and A.~A.~Starobinsky,
  ``A New Universal Local Feature in the Inflationary Perturbation Spectrum,''
  Phys.\ Rev.\ D {\bf 77}, 023514 (2008)
  [arXiv:0711.1585 [astro-ph]].

\bibitem{Joy:2008qd}
  M.~Joy, A.~Shafieloo, V.~Sahni and A.~A.~Starobinsky,
  ``Is a step in the primordial spectral index favored by CMB data ?,''
  JCAP {\bf 0906}, 028 (2009)
  [arXiv:0807.3334 [astro-ph]].

\bibitem{Brandenberger:1990wu}
  J.~H.~Kung and R.~H.~Brandenberger,
  ``Chaotic Inflation as an Attractor in Initial Condition Space,''
  Phys.\ Rev.\ D {\bf 42}, 1008 (1990).

\bibitem{Brandenberger:2003vk}
  R.~H.~Brandenberger,
  ``Lectures on the theory of cosmological perturbations,''
  Lect.\ Notes Phys.\  {\bf 646}, 127 (2004)
  [hep-th/0306071].

\bibitem{Mortonson:2010er}
  M.~J.~Mortonson, H.~V.~Peiris and R.~Easther,
  ``Bayesian Analysis of Inflation: Parameter Estimation for Single Field Models,''
  Phys.\ Rev.\ D {\bf 83}, 043505 (2011)
  [arXiv:1007.4205 [astro-ph.CO]].

\bibitem{Easther:2011yq}
  R.~Easther and H.~V.~Peiris,
  ``Bayesian Analysis of Inflation II: Model Selection and Constraints on Reheating,''
  Phys.\ Rev.\ D {\bf 85}, 103533 (2012)
  [arXiv:1112.0326 [astro-ph.CO]].

\bibitem{Price:2014xpa}
  L.~C.~Price, J.~Frazer, J.~Xu, H.~V.~Peiris and R.~Easther,
  ``MultiModeCode: An efficient numerical solver for multifield inflation,''
  JCAP {\bf 1503}, no. 03, 005 (2015)
  [arXiv:1410.0685 [astro-ph.CO]].

\bibitem{Mukhanov:1990me}
  V.~F.~Mukhanov, H.~A.~Feldman and R.~H.~Brandenberger,
  ``Theory of cosmological perturbations. Part 1. Classical perturbations. Part 2. Quantum theory of perturbations. Part 3. Extensions,''
  Phys.\ Rept.\  {\bf 215}, 203 (1992).

\bibitem{Lewis:1999bs}
  A.~Lewis, A.~Challinor and A.~Lasenby,
  ``Efficient computation of CMB anisotropies in closed FRW models,''
  Astrophys.\ J.\  {\bf 538}, 473 (2000)
  [astro-ph/9911177].

\bibitem{Lewis:2002ah}
  A.~Lewis and S.~Bridle,
  ``Cosmological parameters from CMB and other data: A Monte Carlo approach,''
  Phys.\ Rev.\ D {\bf 66}, 103511 (2002)
  [astro-ph/0205436].

\bibitem{Gelman:1992zz}
  A.~Gelman and D.~B.~Rubin,
  ``Inference from Iterative Simulation Using Multiple Sequences,''
  Statist.\ Sci.\  {\bf 7}, 457 (1992).

\bibitem{Lewis:2013hha}
  A.~Lewis,
  ``Efficient sampling of fast and slow cosmological parameters,''
  Phys.\ Rev.\ D {\bf 87}, no. 10, 103529 (2013)
  [arXiv:1304.4473 [astro-ph.CO]].

\bibitem{Contaldi:2014zua}
  C.~R.~Contaldi, M.~Peloso and L.~Sorbo,
  ``Suppressing the impact of a high tensor-to-scalar ratio on the temperature anisotropies,''
  JCAP {\bf 1407}, 014 (2014)
  [arXiv:1403.4596 [astro-ph.CO]].

\bibitem{Ashoorioon:2014nta}
  A.~Ashoorioon, K.~Dimopoulos, M.~M.~Sheikh-Jabbari and G.~Shiu,
  ``Non-Bunch-Davis initial state reconciles chaotic models with BICEP and Planck,''
  Phys.\ Lett.\ B {\bf 737}, 98 (2014)
  [arXiv:1403.6099 [hep-th]].

\bibitem{Feng:2003mk}
  B.~Feng, M.~Li, R.~J.~Zhang and X.~Zhang,
  ``An inflation model with large variations in spectral index,''
  Phys.\ Rev.\ D {\bf 68}, 103511 (2003)
  [astro-ph/0302479].

\bibitem{Wan:2014fra}
  Y.~Wan, S.~Li, M.~Li, T.~Qiu, Y.~Cai and X.~Zhang,
  ``Single field inflation with modulated potential in light of the Planck and BICEP2,''
  Phys.\ Rev.\ D {\bf 90}, 023537 (2014)
  [arXiv:1405.2784 [astro-ph.CO]].

\bibitem{Miranda:2014wga}
  V.~Miranda, W.~Hu and P.~Adshead,
  ``Steps to Reconcile Inflationary Tensor and Scalar Spectra,''
  Phys.\ Rev.\ D {\bf 89}, no. 10, 101302 (2014)
  [arXiv:1403.5231 [astro-ph.CO]].

\bibitem{Firouzjahi:2014fda}
  H.~Firouzjahi and M.~H.~Namjoo,
  ``Jump in fluid properties of inflationary universe to reconcile scalar and tensor spectra,''
  Phys.\ Rev.\ D {\bf 90}, 063525 (2014)
  [arXiv:1404.2589 [astro-ph.CO]].

\bibitem{Piao:2003zm}
  Y.~S.~Piao, B.~Feng and X.~m.~Zhang,
  ``Suppressing CMB quadrupole with a bounce from contracting phase to inflation,''
  Phys.\ Rev.\ D {\bf 69}, 103520 (2004)
  [hep-th/0310206];\\

\bibitem{Xia:2014tda}
  J.~Q.~Xia, Y.~F.~Cai, H.~Li and X.~Zhang,
  ``Evidence for bouncing evolution before inflation after BICEP2,''
  Phys.\ Rev.\ Lett.\  {\bf 112}, 251301 (2014)
  [arXiv:1403.7623 [astro-ph.CO]].

\bibitem{Cai:2014hja}
  Y.~F.~Cai and Y.~Wang,
  ``Testing quantum gravity effects with latest CMB observations,''
  Phys.\ Lett.\ B {\bf 735}, 108 (2014)
  [arXiv:1404.6672 [astro-ph.CO]].

\bibitem{Cai:2014bea}
  Y.~F.~Cai,
  ``Exploring Bouncing Cosmologies with Cosmological Surveys,''
  Sci.\ China Phys.\ Mech.\ Astron.\  {\bf 57}, 1414 (2014)
  [arXiv:1405.1369 [hep-th]].

\bibitem{Hamann:2007pa}
  J.~Hamann, L.~Covi, A.~Melchiorri and A.~Slosar,
  ``New Constraints on Oscillations in the Primordial Spectrum of Inflationary Perturbations,''
  Phys.\ Rev.\ D {\bf 76}, 023503 (2007)
  [astro-ph/0701380].

\bibitem{Jain:2008dw}
  R.~K.~Jain, P.~Chingangbam, J.~O.~Gong, L.~Sriramkumar and T.~Souradeep,
  ``Punctuated inflation and the low CMB multipoles,''
  JCAP {\bf 0901}, 009 (2009)
  [arXiv:0809.3915 [astro-ph]].

\bibitem{Jain:2009pm}
  R.~K.~Jain, P.~Chingangbam, L.~Sriramkumar and T.~Souradeep,
  ``The tensor-to-scalar ratio in punctuated inflation,''
  Phys.\ Rev.\ D {\bf 82}, 023509 (2010)
  [arXiv:0904.2518 [astro-ph.CO]].

\bibitem{Mortonson:2009qv}
  M.~J.~Mortonson, C.~Dvorkin, H.~V.~Peiris and W.~Hu,
  ``CMB polarization features from inflation versus reionization,''
  Phys.\ Rev.\ D {\bf 79}, 103519 (2009)
  [arXiv:0903.4920 [astro-ph.CO]].

\bibitem{Hazra:2010ve}
  D.~K.~Hazra, M.~Aich, R.~K.~Jain, L.~Sriramkumar and T.~Souradeep,
  ``Primordial features due to a step in the inflaton potential,''
  JCAP {\bf 1010}, 008 (2010)
  [arXiv:1005.2175 [astro-ph.CO]].

\bibitem{Hazra:2014jka}
  D.~K.~Hazra, A.~Shafieloo, G.~F.~Smoot and A.~A.~Starobinsky,
  ``Inflation with Whip-Shaped Suppressed Scalar Power Spectra,''
  Phys.\ Rev.\ Lett.\  {\bf 113}, no. 7, 071301 (2014)
  [arXiv:1404.0360 [astro-ph.CO]].

\bibitem{Hazra:2014goa}
  D.~K.~Hazra, A.~Shafieloo, G.~F.~Smoot and A.~A.~Starobinsky,
  ``Wiggly Whipped Inflation,''
  JCAP {\bf 1408}, 048 (2014)
  [arXiv:1405.2012 [astro-ph.CO]].

\bibitem{Hazra:2013nca}
  D.~K.~Hazra, A.~Shafieloo and G.~F.~Smoot,
  ``Reconstruction of broad features in the primordial spectrum and inflaton potential from Planck,''
  JCAP {\bf 1312}, 035 (2013)
  [arXiv:1310.3038 [astro-ph.CO]].

\bibitem{Abazajian:2014tqa}
  K.~N.~Abazajian, G.~Aslanyan, R.~Easther and L.~C.~Price,
  ``The Knotted Sky II: Does BICEP2 require a nontrivial primordial power spectrum?,''
  JCAP {\bf 1408}, 053 (2014)
  [arXiv:1403.5922 [astro-ph.CO]].

\bibitem{Hu:2014aua}
  B.~Hu, J.~W.~Hu, Z.~K.~Guo and R.~G.~Cai,
  ``Reconstruction of the primordial power spectra with Planck and BICEP2,''
  Phys.\ Rev.\ D {\bf 90}, 023544 (2014)
  [arXiv:1404.3690 [astro-ph.CO]].

\bibitem{Ashoorioon:2006wc}
  A.~Ashoorioon and A.~Krause,
  ``Power Spectrum and Signatures for Cascade Inflation,''
  hep-th/0607001.

\bibitem{Ashoorioon:2008qr}
  A.~Ashoorioon, A.~Krause and K.~Turzynski,
  ``Energy Transfer in Multi Field Inflation and Cosmological Perturbations,''
  JCAP {\bf 0902}, 014 (2009)
  [arXiv:0810.4660 [hep-th]].

\bibitem{Ashoorioon:2014yua}
  A.~Ashoorioon, C.~van de Bruck, P.~Millington and S.~Vu,
  ``Effect of transitions in the Planck mass during inflation on primordial power spectra,''
  Phys.\ Rev.\ D {\bf 90}, no. 10, 103515 (2014)
  [arXiv:1406.5466 [astro-ph.CO]].

\bibitem{Ballesteros:2014yva}
  G.~Ballesteros and J.~A.~Casas,
  ``Large tensor-to-scalar ratio and running of the scalar spectral index with Instep Inflation,''
  Phys.\ Rev.\ D {\bf 91}, 043502 (2015)
  [arXiv:1406.3342 [astro-ph.CO]].

\bibitem{Cai:2015dta}
  Y.~Cai, Y.~T.~Wang and Y.~S.~Piao,
  ``Oscillating modulation to B-mode polarization from varying propagating speed of primordial gravitational waves,''
  Phys.\ Rev.\ D {\bf 91}, 103001 (2015)
  [arXiv:1501.06345 [astro-ph.CO]].

\bibitem{Cai:2015xla}
  Y.~F.~Cai, E.~G.~M.~Ferreira, B.~Hu and J.~Quintin,
  ``Searching for features of a string-inspired inflationary model with cosmological observations,''
  Phys.\ Rev.\ D {\bf 92}, no. 12, 121303 (2015)
  [arXiv:1507.05619 [astro-ph.CO]].

\bibitem{Flauger:2014ana}
  R.~Flauger, L.~McAllister, E.~Silverstein and A.~Westphal,
  ``Drifting Oscillations in Axion Monodromy,''
  arXiv:1412.1814 [hep-th].

\end{thebibliography}
\end{document}